\documentclass[twocolumn]{aastex631}

\usepackage{graphicx}	
\usepackage{amsmath}	
\usepackage{amssymb}	

\usepackage{amsmath}
\usepackage{bm}
\usepackage{newtxtext,newtxmath}

\usepackage{graphicx}	
\usepackage{grffile}
\usepackage{amsmath}	
\usepackage{amssymb}	

\usepackage{epsfig,amsmath,natbib}
\usepackage{color,subfigure}
\usepackage{xcolor}

\usepackage{mathrsfs,amssymb,amstext}
\usepackage{ulem}
\usepackage{url}
\usepackage{bm}

\usepackage[T1]{fontenc}

\DeclareRobustCommand{\VAN}[3]{#2}
\let\VANthebibliography\thebibliography
\def\thebibliography{\DeclareRobustCommand{\VAN}[3]{##3}\VANthebibliography}

\def\mean#1{\left< #1 \right>}

\begin{document}

\title{On Measuring the 21 cm Global Spectrum of the Cosmic Dawn with an Interferometer Array }

\correspondingauthor{Bin Yue; Xuelei Chen}
\email{yuebin@nao.cas.cn; xuelei@cosmology.bao.ac.cn}

\author{Xin Zhang}
\affiliation{National Astronomical Observatories, Chinese Academy of Sciences, 20A, Datun Road, Chaoyang District, Beijing 100101, China}
\affiliation{School of Astronomy and Space Science, University of Chinese Academy of Sciences, Beijing 100049, China}

\author{Bin Yue}
\affiliation{National Astronomical Observatories, Chinese Academy of Sciences, 20A, Datun Road, Chaoyang District, Beijing 100101, China}

\author{Yuan Shi}
\affiliation{National Astronomical Observatories, Chinese Academy of Sciences, 20A, Datun Road, Chaoyang District, Beijing 100101, China}
\affiliation{School of Astronomy and Space Science, University of Chinese Academy of Sciences, Beijing 100049, China}

\author{Fengquan Wu}
\affiliation{National Astronomical Observatories, Chinese Academy of Sciences, 20A, Datun Road, Chaoyang District, Beijing 100101, China}

\author{Xuelei Chen}
\affiliation{National Astronomical Observatories, Chinese Academy of Sciences, 20A, Datun Road, Chaoyang District, Beijing 100101, China}
\affiliation{School of Astronomy and Space Science, University of Chinese Academy of Sciences, Beijing 100049, China}
\affiliation{Department of Physics, College of Sciences, Northeastern University, Shenyang 110819, China}
\affiliation{Center of High Energy Physics, Peking University, Beijing 100871, China}

\shorttitle{21 cm Global Spectrum from Interferometer}
\shortauthors{Zhang et al.}

\begin{abstract}
We theoretically investigate  the recovery of global spectrum (monopole)  from visibilities (cross-correlation only) measured by the interferometer array and the feasibility of extracting 21 cm signal of cosmic dawn.
In our approach, the global spectrum is obtained by solving the monopole and higher-order components simultaneously from the visibilities measured with up to thousands of baselines. Using this algorithm, the monopole of both foreground and the 21 cm signal
can be correctly recovered in a broad range  of conditions. We find that a 3D baseline distribution can have much better performance than a 2D (planar) baseline distribution, particularly when there is a lack of shorter baselines. We simulate for ground-based 2D and 3D array configurations, and a cross-shaped space array located at the Sun-Earth L2 point that can form 3D baselines through orbital precession. In all simulations we obtain good recovered global spectrum, and successfully extract the 21 cm signal from it, with reasonable number of antennas and observation time.

\end{abstract}


\section{Introduction}

The 21 cm emission line is produced by transition between the two hyper-fine energy levels of the neutral hydrogen atom (H). It is the most  promising tool that can directly and efficiently detect multiple cosmic stages from the dark ages to reionization  (e.g. \citealt{Furlanetto2006PhR,Furlanetto2006MNRAS.371..867F,Furlanetto2006ApJ...652..849F,Chen_2008,Yue2009,Xu2009ApJ,Xu2011MNRAS}). The distribution of neutral hydrogen atoms at the two energy levels is determined by the balance between the absorption/re-emission of CMB photons by H atom, the H$-$H and H$-e^-$ collisions, and the H$-$Ly$\alpha$ scattering. It is described by the spin temperature $T_s$. If $T_s$ is smaller than the CMB temperature $T_{\rm CMB}$,  the 21 cm  signal is absorption feature on the CMB spectrum.

After the intergalactic medium (IGM) decouples from the CMB and before the X-ray heating works efficiently ($150\gtrsim z\gtrsim15$, e.g. \citealt{Pritchard2012RPPh,Barkana2001PhR}), the kinetic temperature of the IGM, $T_k$, is always smaller than the CMB temperature.  Two mechanisms make the $T_s$ couple tightly to the $T_k$: the first one is H$-$H and H$-e^-$ collisions, it generates an absorption trough $\sim$50 mK at the dark ages around $z\sim70$ \citep{Loeb2004PRL}; the second one is through the H$-$Ly$\alpha$ scattering (Wouthuysen-Field effect, \citealt{Wouthuysen1952,Field1958}), it generates a much stronger absorption trough at cosmic dawn around $z\sim20$ \citep{Chen2004,Hirata2006MNRAS,Hirata2009PRD}. 
The amplitude of the absorption trough at cosmic dawn can reach up to $\sim$200 mK \citep{Cohen2017,Xu2018ApJ,Villanueva-Domingo2020PRD,Xu2021ApJ}, which is the most promising feature for 21 cm global spectrum detection. Such a signal would provide rich information about the cosmic dawn  (e.g. \citealt{Furlanetto2006Ly,Madau2018,Fialkov2013,Cohen2020,Mittal2021,Mebane2020MNRAS,Mirocha2018MNRAS,Monsalve2019ApJ}).

A number of experiments are dedicated to measure the global spectrum for the cosmic dawn and epoch of reionization, such as the EDGES \citep{2008ApJ...676....1B,Bowman2010Nature}; SARAS  \citep{2013ExA....36..319P,2018ExA....45..269S,Singh2018ApJ,SARAS3}; SCI-HI \citep{SCI-HI2014ApJ}; BIGHORNS \citep{BIGHORNS2015PASA}; LEDA \citep{Bernardi2016MNRAS,2018MNRAS.478.4193P}; PRI$^{Z}$M \citep{2019JAI.....850004P}; ASSASSIN \citep{ASSASSIN2020MNRAS}; 
REACH \citep{REACH2022NatAs,Anstey_2022MNRAS,Cumner_2022JAI};
and so on.  Particularly, the EDGES experiment has detected an absorption signal at $\sim$ 80 MHz, with an amplitude $\sim 550$ mK \citep{Bowman2018Nature}. 
The corresponding redshift for the  21 cm line at this frequency is indeed where we expect for the cosmic dawn, however the amplitude is much stronger than even the most optimistic model in standard cosmology. If it is true, then it may imply the existence of exotic new physics, which could provide additional cooling for the gas  \citep{Barkana2018Nature} or an extra radio background during the cosmic dawn \citep{Feng2018ApJ,Ewall-Wice2020MNRAS,Fialkov2019MNRAS, Ewall-Wice2018ApJ}. 
Moreover, the width of the absorption trough raises a question of star formation in high-$z$  galaxies.
The observed width is narrow, $\sim20$ MHz. It 
 implies that the Ly$\alpha$ coupling and X-ray heating must start to work efficiently  at $z\sim20$ and $z\sim15$. If the relevant photons are provided by  high-$z$ galaxies, their UV  luminosity functions must have steep faint-ends. Star formation in halos below $\sim10^8-10^{10}~M_\odot$ should be much more efficient than expected  in the general model. See an intensive study in \cite{Mirocha2019MNRAS}.  
However, the required precision to measure such a 21 cm signal is so high, it is easy to be affected by even small systematic effects. Recent  measurement by SARAS-3 \citep{Singh2021} claims no-detection of 21 cm absorption signal at the relevant redshift. Measurements by alternative instruments, or even based on alternative principles, are earnestly needed to resolve the issue. 

A radio interferometer measures the cross-correlation of the signal sensed by a pair of sensors, as such it is easier to distinguish the signal and noise, and less susceptible to calibration error. As the interferometer measures the spatial variations of sky radiation, one might think that it is irrelevant to the global signal measurement.  However, \citet{Liu2013} suggested that by measuring the sky spectrum with telescope of finer angular resolution, the global 21 cm signal and foreground can be better separated. 
The angular structure information is also used to extract the 21 cm global from the foreground in some recently developed algorithms \citep{Rapetti2020ApJ,Tauscher2020ApJ,REACH2022NatAs}. 
\citet{2014arXiv1406.2585M} proposed that by placing a beam-splitter (a vertical resistive sheet which can both transmit and reflect part of the incident radio wave) between two antennas placed at almost zero distance, a zero-spacing interferometer can be formed, and it can measure the mean sky brightness. \citet{Presley2015}
noted that although the cross-correlation of the interferometer is only sensitive to the variations of sky intensity, the monopole component does contribute to the visibility when  the non-constant primary beam of the antenna is taken into account, or if the modulation of lunar occultation is considered \citep{2015MNRAS.450.2291V}. Suppose the sky brightness is uniform, or if baselines are sufficiently short ($\ll \lambda$ ), the monopole can dominate over other higher orders components in the measured visibility, then it should be possible to measure this global signal with a regular interferometer array, or even with an interferometer of aperture array elements \citep{2015ApJ...815...88S}. \citet{Presley2015} concluded that the measurement of the global spectrum relies on the short baselines, and tightly packed array is required. As small physical size of antennas allow them to be packed closer, while larger physical size antennas provides more modulation with primary beam, an intermediate size is preferred, so that the full width half-max (FWHM) of the primary beam is about $40^\circ$.
In most of these works the cross-talks between different elements of the interferometer are neglected. \citet{2016ApJ...826..116V}, however, pointed out that if short baselines are employed the cross-talk is inevitable, and it will generate systematic bias. \citet{ASSASSIN2020MNRAS}  moved forward to implement the interferometric global spectrum measurement with actural arrays.
They simulated recovering global sky temperature and extracting 21 cm signal for the SKA EDA-2 array configuration, and applied their  methods to the real data measured by the EDA-2. A dedicated array named SITARA has also been proposed, with model of cross-talks \citep{Thekkeppattu2022}.

When working on the synthesis imaging problem of the Discovering the Sky at the Longest wavelength (DSL) project \citep{Chen2019arXiv,Chen2020}, which is a lunar orbit interferometer array, 
some of us found that the full sky map can be reconstructed very well from the interferometric data \citep{Huang2018,Shi2022}. Although we were not particularly seeking to measure the global spectrum using interferometry there (in the DSL project, the global spectrum is to be measured with single antenna, see \citealt{Shi2022_global}), in one simulation where we assumed uniform primary beam, the full sky map is well-recovered, suggesting that even without the modulation of the primary beam, the monopole is still recoverable from interferometric data. 
The reason is: although in the 2D approximation, each baseline corresponds to a specific spatial frequency according to the Van Cittert-Zernike theorem \citep{Thompson2017interferometry}, when the full 3D sky is considered, each baseline does not correspond to a single spherical harmonic mode, and the monopole actually also contribute to each visibility, thus its information can in principle be extracted, see the  Appendix A of \cite{2015MNRAS.450.2291V}.

In this paper, we investigate the feasibility of recovering the monopole component of the sky temperature along with higher order components simultaneously from the visibilities measured by many baselines of an interferometer array.  The advantage of such a solution is that the angular-response correction will not rely on a precise sky model. Moreover, it is important to know whether the fluctuations (induced by instrumental noise and baseline distribution) of the recovered global spectrum  configurations can be well controlled for realistic array, so that it will not confuse the 21 cm signal. We  particularly focus on the influence of noise, beam and baseline distribution. This is the motivation of this work. 

The paper is organized as follows: in Sec. \ref{sec:algorithm} we introduce our  algorithm and test its feasibility; in Sec. \ref{sec:recover-21cm} we present the simulation results for ground-based and space arrays, and some discussions. These are our major results.
Finally  we reach our conclusions in Sec. \ref{sec:conclusion}.

\section{The algorithm}\label{sec:algorithm}

\begin{figure}
\centering{
\subfigure{\includegraphics[width=0.45\textwidth]{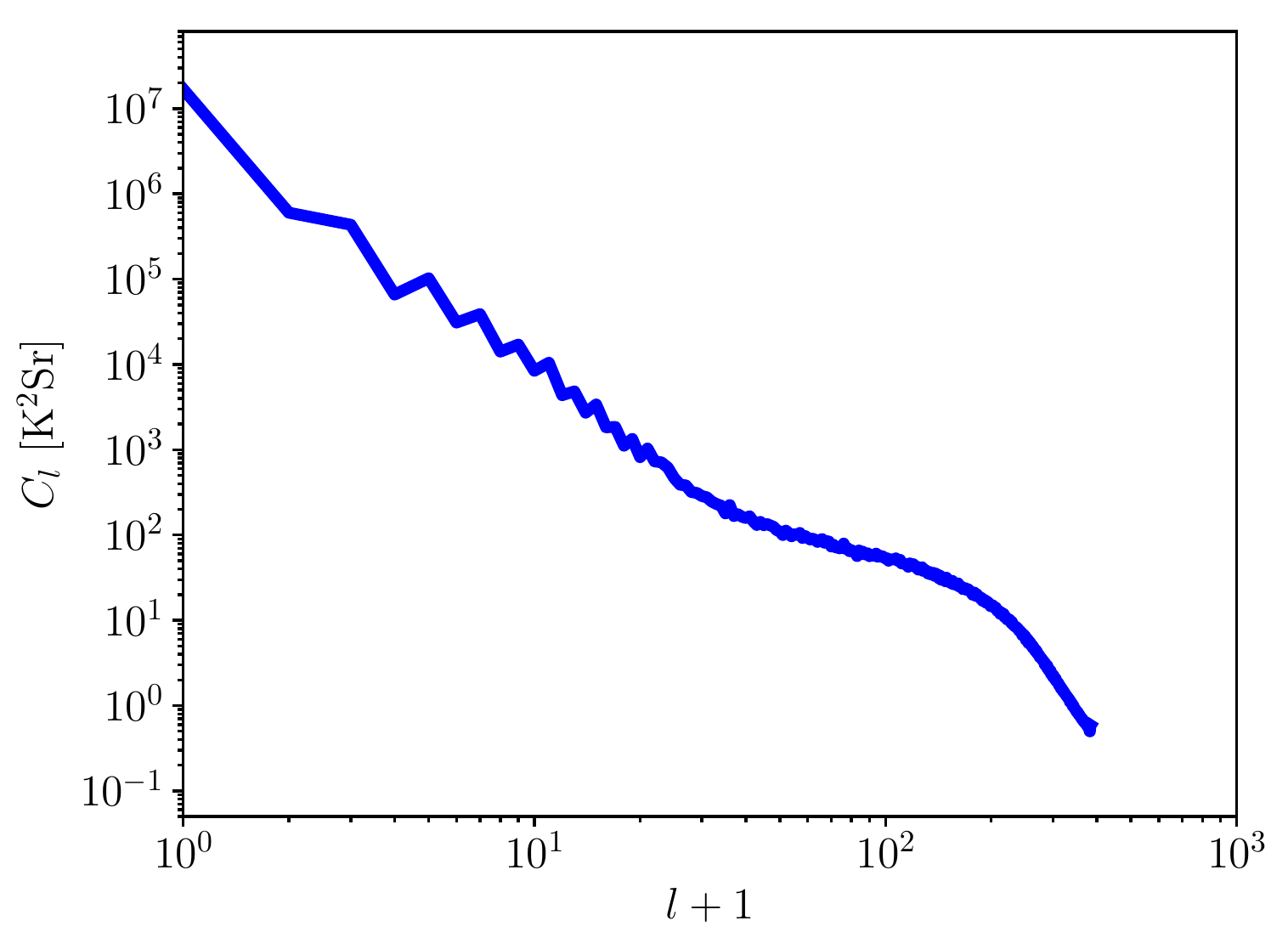}}
\caption{The angular power spectrum of the adopted sky model map at 100 MHz.
}
\label{fig:aps}
}
\end{figure}

The interferometric visibility of a pair of antennas with baseline $\boldsymbol{b}$ at frequency $\nu$ and wavelength $\lambda$ is given by 
\begin{equation}
V_\nu(\boldsymbol{b}, \hat{\boldsymbol{n}}_0)= \int d \Omega(\hat{\boldsymbol{n}}) B_\nu(\hat{\boldsymbol{n}}, \hat{\boldsymbol{n}}_0) T_\nu(\hat{\boldsymbol{n}}) e^{-2 \pi i \frac{\boldsymbol{b}}{\lambda} \cdot \hat{\boldsymbol{n}}},
\label{eq:V}
\end{equation}
where $T_\nu(\hat{\boldsymbol{n}})$ is the sky temperature at the direction $\hat{\boldsymbol{n}}$, $B_\nu$ is the primary beam and $\hat{\boldsymbol{n}}_0$ is the beam center (the direction where beam is maximal). Here we have neglected the noise, contribution of radio frequency interference (RFI), ground pick up, cross-coupling between the interferometer array elements and other non-linear effects.  
We adopt normalization that $B_\nu(\hat{\boldsymbol{n}}_0)=1$.  
Throughout this paper we use the {\tt ULSA} sky model with direction-dependent spectrum indices  \citep{Cong2021} as our input sky map. We degrade the map to NSIDE=128, corresponding to angular resolution $\sim 0.5^\circ$. 
Throughout this paper, we only consider the baselines shorter than 10$\times$wavelength. These baselines are only sensitive to sky temperature anisotropy $\gtrsim 5^\circ$. So the angular resolution of the sky map is more than enough for our investigation.
We show the angular power spectrum of the map at 100 MHz in Fig. \ref{fig:aps}.
Expanding the sky temperature $T_\nu$ by spherical harmonic functions,
\begin{equation}
T_\nu(\hat{\boldsymbol{n}}) = \sum_{l=0}^{\infty}\sum_{m=-l}^{l} a_l^m Y_l^m(\hat{\boldsymbol{n}} ),
\label{eq:a_lm}
\end{equation}
the visibility can then be expanded, up to $l_{\rm max}$, as \citep{Presley2015}
\begin{equation}
V_\nu(\boldsymbol{b}, \hat{\boldsymbol{n}}_0)\approx \sum_{l=0}^{l_{\rm max}}\sum_{m=-l}^{l}a_l^m\left(\int d \Omega(\hat{\boldsymbol{n}}) B_\nu(\hat{\boldsymbol{n}}, \hat{\boldsymbol{n}}_0) Y_l^m(\hat{\boldsymbol{n}} )  e^{-2 \pi i \frac{\boldsymbol{b}}{\lambda} \cdot \hat{\boldsymbol{n}}}\right).
\label{eq:V2}
\end{equation}

We may write the linear equations of many baselines in matrix form, with noise $\boldsymbol{V}_{\rm N}$, 
\begin{equation}
\boldsymbol{V}=\boldsymbol{Q} \boldsymbol{a}+\boldsymbol{V}_{\rm N},
\label{eq:V3}
\end{equation}  
where for the $j$-th baseline $\boldsymbol{b}_j$ 
\begin{equation}
Q_{l,j}^m=\left(\int d \Omega(\hat{\boldsymbol{n}}) B_\nu(\hat{\boldsymbol{n}}, \hat{\boldsymbol{n}}_0) Y_l^m(\hat{\boldsymbol{n}} )  e^{-2 \pi i \frac{\boldsymbol{b}_j}{\lambda} \cdot \hat{\boldsymbol{n}}}\right),
\label{eq:Q_jlm}
\end{equation}
and the $\boldsymbol{Q}$ matrix has a total of $(l_{\rm max}+1)^2$ columns. 

Given visibility data from a sufficient number of baselines, we should be able to solve these equations and recover these multipole moments of sky temperature.
However, we must take into account another constraint: the sky temperature $T_\nu(\hat{\boldsymbol{n}})$ should be real numbers. In this case there is $a_{l}^{-m}=(-1)^m(a_{l}^{m})^*$. To force the solution to satisfy this symmetry requirement, we separate the $\boldsymbol{V}$, $\boldsymbol{Q}$ and $\boldsymbol{a}$ into real and imaginary parts, and build a new equation whose unknowns are only those $a_l^m$s with $m\ge 0$, 
\begin{equation}
\tilde{\boldsymbol{V}}=\tilde{\boldsymbol{Q}}\tilde{\boldsymbol{a}},
\label{eq:V_tilde}
\end{equation}
where  $\tilde{\boldsymbol{V}}$ is a column-vector that is composed of the real and imaginary parts of the vector $\boldsymbol{V}$, $\tilde{{\boldsymbol{V}}}=[ ...{\rm Re}(V_j)..., ...{\rm Im}(V_j)...]^{\rm T}$.
$\tilde{\boldsymbol{a}}$ is composed of the real and imaginary parts of the vector $\boldsymbol{a}(m\ge0)$, $\tilde{a}=[...{\rm Re}(a_l^m)...,...{\rm Im}(a_l^m)...]^{\rm T}$.
\begin{equation}
\tilde{\boldsymbol{Q}}=\left[
\begin{array}{cc}
\boldsymbol{A}, \boldsymbol{B} \\
\boldsymbol{C}, \boldsymbol{D} 
\end{array}
\right],
\end{equation}
where
\begin{align}
A_{l,j}^m&={\rm Re}(Q_{l,j}^m)+s(m)(-1)^m{\rm Re}(Q_{l,j}^{-m}) \nonumber \\
B_{l,j}^{m}&=-{\rm Im}(Q_{l,j}^m)+s(m)(-1)^m{\rm Im}(Q_{l,j}^{-m}) \nonumber \\
C_{l,j}^m&={\rm Im}(Q_{l,j}^m)+s(m)(-1)^m{\rm Im}(Q_{l,j}^{-m}) \nonumber \\
D_{l,j}^m&={\rm Re}(Q_{l,j}^m)-s(m)(-1)^m{\rm Re}(Q_{l,j}^{-m}), 
\label{eq:ABCD}
\end{align}
where
\begin{equation}
s(m)=
\begin{cases}
0~~{\rm when}~m=0 \\
1~~{\rm otherwise}.
\end{cases}
\end{equation} 
Suppose we have $N_L$ baselines and solve for multipole coefficients up to $l_{\rm max}$, then the matrix  $\tilde{\boldsymbol{Q}}$ has $2N_L$ rows and 
$(l_{\rm max}+1)(l_{\rm max}+2)$
columns. 
The solution, or estimator of the sky temperature harmonic coefficients are given by 
\begin{equation}
\hat{\tilde{\boldsymbol{a}}}=\tilde{\boldsymbol{Q}}^{-1} \tilde{\boldsymbol{V}},
\label{eq:a_hat}
\end{equation}
where $\tilde{\boldsymbol{Q}}^{-1}$ denotes either the inverse matrix of $\tilde{\boldsymbol{Q}}$, or a pseudo-inverse (e.g. the Moore-Penrose pseudo-inverse) if the inverse matrix does not exist.  
The recovered global sky temperature is then given by  
$\hat{T}_{\rm sky}(\nu)=\hat{\tilde{a}}_0Y_{00}=\hat{\tilde{a}}_0/\sqrt{4\pi}$, where $\hat{\tilde{a}}_0$ is the first element of $\hat{\tilde{\boldsymbol{a}}}$. 
 
We first investigate whether the above algorithm is feasible in an ideal case where we have no noise and an isotropic beam for each antenna, $B_\nu\equiv1$, we also assume here the whole sky is visible, i.e. there is no blocking by the ground, and the cross-couplings between the antennas are negligible.
In this case Eq.~(\ref{eq:Q_jlm}) can be simplified as 
\begin{eqnarray}
Q_{l,j}^m =(-1)^l4 \pi   i^{l} \mathcal{J}_{l}(2 \pi b_j/\lambda) Y_{l}^m(\hat{\boldsymbol{b}}_j) 
 \label{eq:Q_jlm_iso}
\end{eqnarray}
by using the relation \\
$$e^{-2 \pi i \frac{\boldsymbol{b}_j}{\lambda} \cdot \hat{\boldsymbol{n}}}=4 \pi \sum_{l^{\prime} m^{\prime}} i^{l^\prime} \mathcal{J}_{l^\prime}(2 \pi b_j/\lambda) Y_{l^{\prime} }^{ m^{\prime}}(-\hat{\boldsymbol{b}}_j) [Y_{l^{\prime} }^{ m^{\prime}}(\hat{\boldsymbol{n}})]^*,
$$
where $\mathcal{J}_{l}$ is the spherical Bessel function of the first kind of order $l$, and $b_j=|\boldsymbol{b}_j|$. 
Note that for the baselines located on the horizontal ground, $\theta_{b_j}=\pi/2$, $Y_l^m(\theta_{b_j},\phi_{b_j})\equiv 0$ when $l-m$ is odd number. In such case almost a half of the columns of the matrix  $\boldsymbol{Q}$ are exactly zero. We compute  $\tilde{\boldsymbol{Q}}^{-1}$ using the {\tt numpy.linalg.pinv} function provided by the {\tt Python NumPy} linear algebra package, and $Y_l^m$ and  $\mathcal{J}_{l}$ by using functions in the {\tt Python scipy} Special Functions package.
To numerically compute $\tilde{\boldsymbol{Q}}^{-1}$, one must specify a cutoff criterion $r_{\rm cut}$. The singular values smaller than  $r_{\rm cut}$ times the largest singular value are discarded. The default cutoff is usually set to the larger of the row number and  column number of the matrix times the machine precision for a 64-bit float number, which is $\mathcal{O}(10^{-12})$ for a matrix with thousands of rows or columns. Here we take this default $r_{\rm cut}$ for our noiseless ideal case, but in the following subsections we will see that in the presence of noise a much larger $r_{\rm cut}$ should be adopted to avoid noise contamination.

\begin{figure}
\centering{
\subfigure{\includegraphics[width=0.45\textwidth]{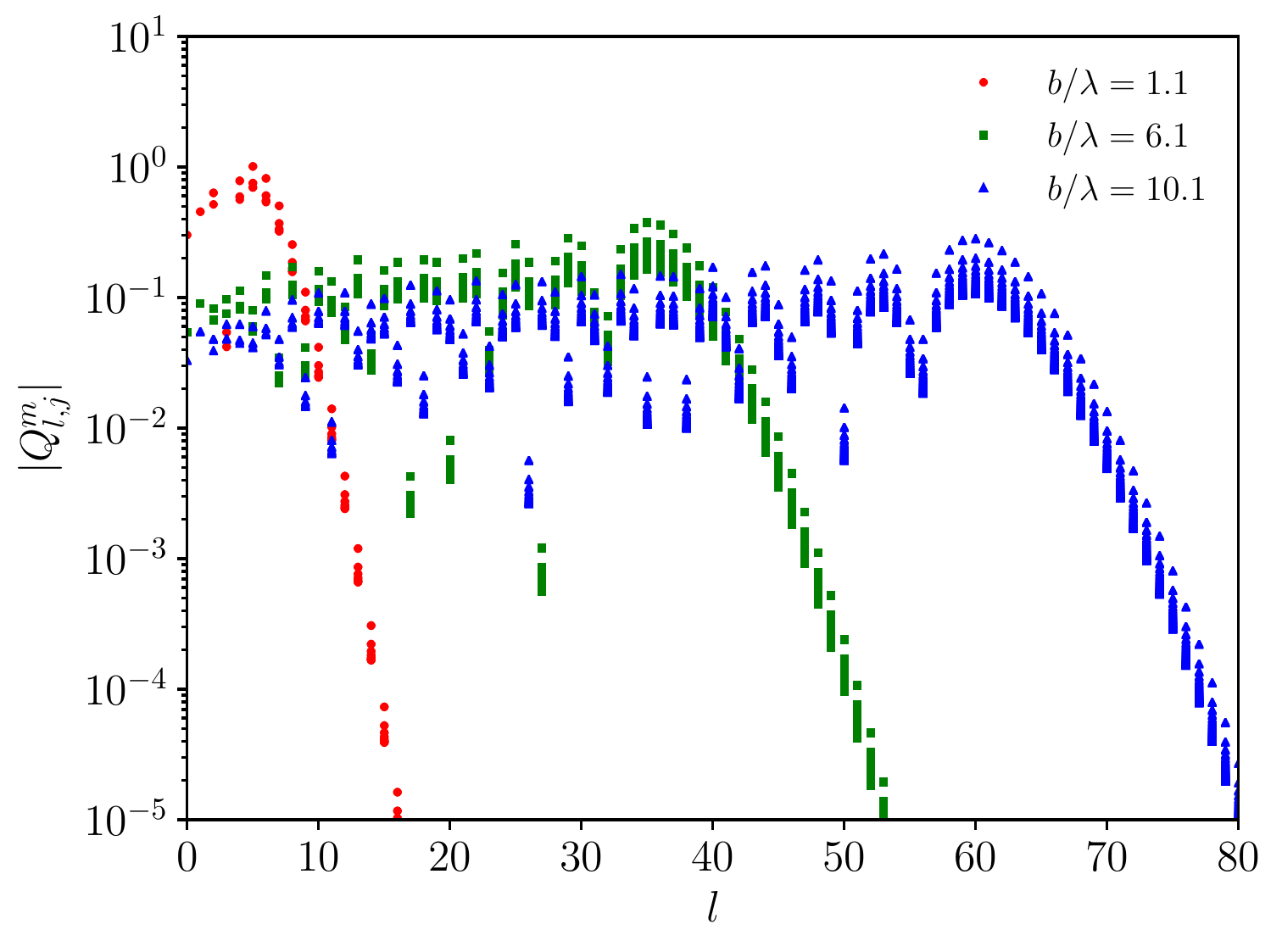}}
\caption{$|Q_{l,j}^m|$ vs. $l$ for different lengths of baselines. Different $m$ values for the same $l$ are all plotted on the same figure. 
}
\label{fig:Q_jlm}
}
\end{figure}

When we place the antennas, the formed baselines will have a range of lengths.
As we are interested primarily about the lowest mode, i.e. the monopole, only the short baselines of a few wavelength are effective. Here we consider the baselines with $b\lesssim10\lambda$.
The interferometric measurement with a given baseline length $b$ is only sensitive to modes up to $l \sim 2\pi b/\lambda$, above which it drops rapidly, as shown in Fig. \ref{fig:Q_jlm} for several baseline lengths.  
Therefore $l_{\rm max}$ must be larger than $ 2\pi b_{\rm max}/\lambda\approx60$.
We take $l_{\rm max}=80$. It is larger than the drop-off scale $l\sim60$ for a safety margin. The  
$|Q_{l}^m| $ at $l>l_{\rm max}$ is smaller than $10^{-4}$ of the $|Q_{l}^m| $ at $l\sim 2\pi b/\lambda$.

We need a sufficiently large $N_L$ so that the least square solution of  Eq. (\ref{eq:a_hat}) has sufficient precision. For the isotropic beam, we generate different number of baselines with lengths set between $b_{\rm min}=\lambda$ and $b_{\rm max}=10\lambda$, randomly distributed on a plane, then solve the visibility equations and obtain the monopole solution. For $N_L \gtrsim 1000$, we   obtained $|\hat{\tilde{a}}_0-{\rm Re}(a_0^0)|/{\rm Re}(a_0^0)\lesssim10^{-8}$.
If we increase (decrease) $b_{\rm max}$, we will need to increase (decrease) $l_{\rm max}$, and then to obtain a good solution we will also need more (less) baselines in the visibility equations correspondingly.  
The minimum $N_L$ (or the number of antennas) depends on the baseline distribution, beam, noise, and the Earth block effects. When these conditions  change we need to check if the $N_L$ is still large enough for giving the converged results, using our sky model as a reference.

\begin{figure}
\centering{
\subfigure{\includegraphics[width=0.45\textwidth]{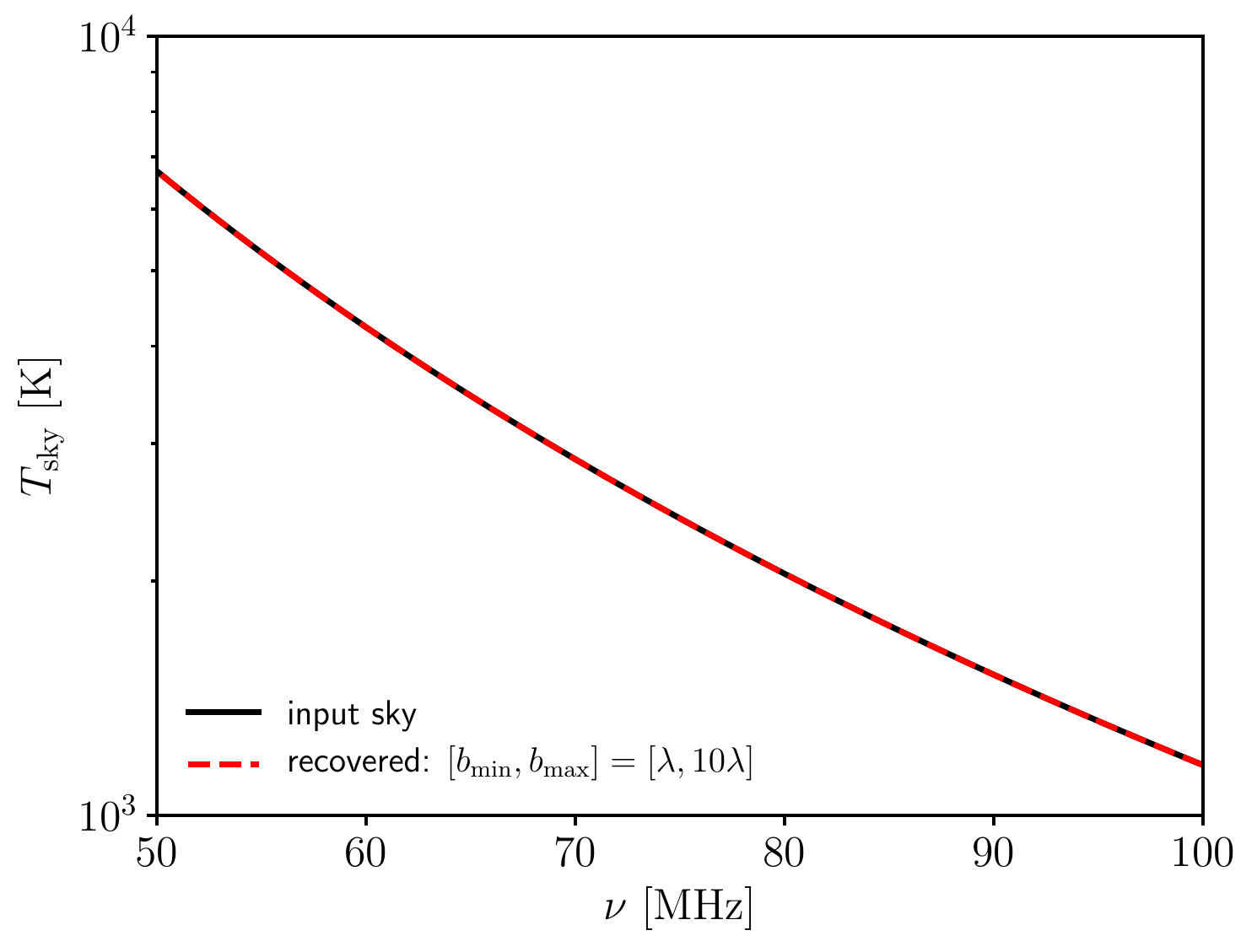}}
\caption{The recovered global sky temperature as a function of frequency, compared with the input sky temperature. We consider baselines with $[b_{\rm min},b_{\rm max}]=[\lambda,10\lambda]$, and take $r_{\rm cut}=2\times10^{-12}$.
}
\label{fig:T_recovered_random2D}
}
\end{figure}

\begin{figure}
\centering{
\subfigure{\includegraphics[width=0.45\textwidth]{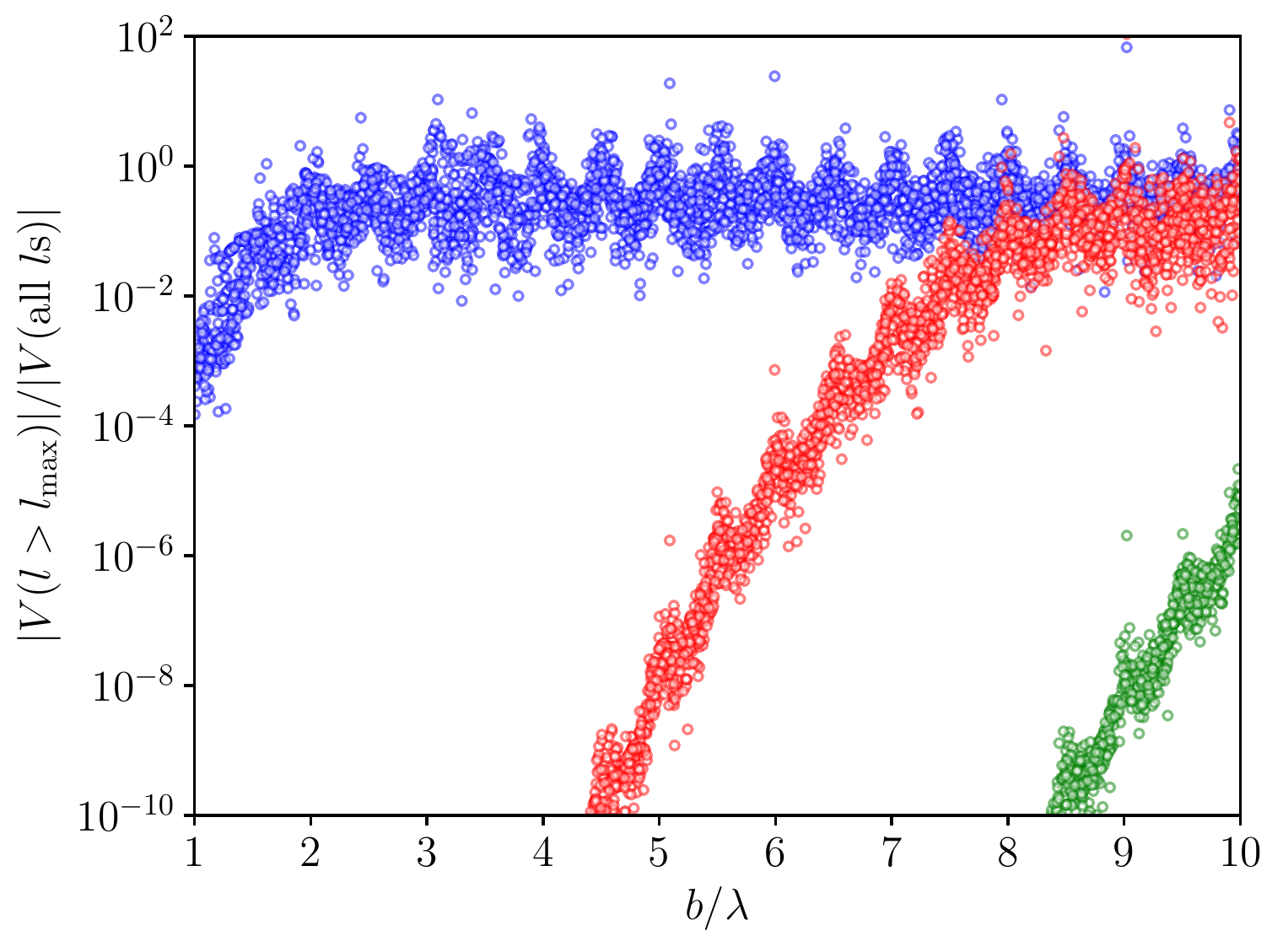}}
\caption{
 The relative residue contribution from $l>l_{\rm max}$ modes to the full visibilities, as a function of the length of the baselines. From top to bottom (colors of blue, red and green), symbol groups correspond to $l_{\rm max}=10$, 50 and 80 respectively.
}
\label{fig:v_vs_b_length}
}
\end{figure}

According to the above considerations, we generate $N_L=4000$  baselines (corresponding to several tens antennas) with lengths between $b_{\rm min}=\lambda$ and $b_{\rm max}=10\lambda$, randomly distributed on a plane.
We neglect couplings between the different baselines. For each baseline we calculate the visibility for the given sky model, then solve $\hat{T}_{\rm sky}(\nu)$ using Eq.(\ref{eq:a_hat}) at each frequency. Fig. \ref{fig:T_recovered_random2D} shows the recovered global spectrum.
In this ideal setup (the beam is isotropic and there is no noise) the global spectrum is recovered perfectly from the interferometric data by Eq. (\ref{eq:a_hat}). The  relative difference between the recovered and the input temperature is only $\sim10^{-9}$. 
In Fig. \ref{fig:v_vs_b_length} we plot the relative contributions from modes with $l > l_{\rm max}$ for all baselines for $l_{\rm max}=10$, 50 and 80 respectively. We see that when $l_{\rm max}=80$, for most baselines the modes with $l \ge l_{\rm max}$ have fractional contribution $\lesssim 10^{-10}$ to visibilities. For some baselines with lengths $\sim10\lambda$, the fractional contribution is $\lesssim10^{-5}$.

\begin{figure}
\centering{
\subfigure{\includegraphics[width=0.45\textwidth]{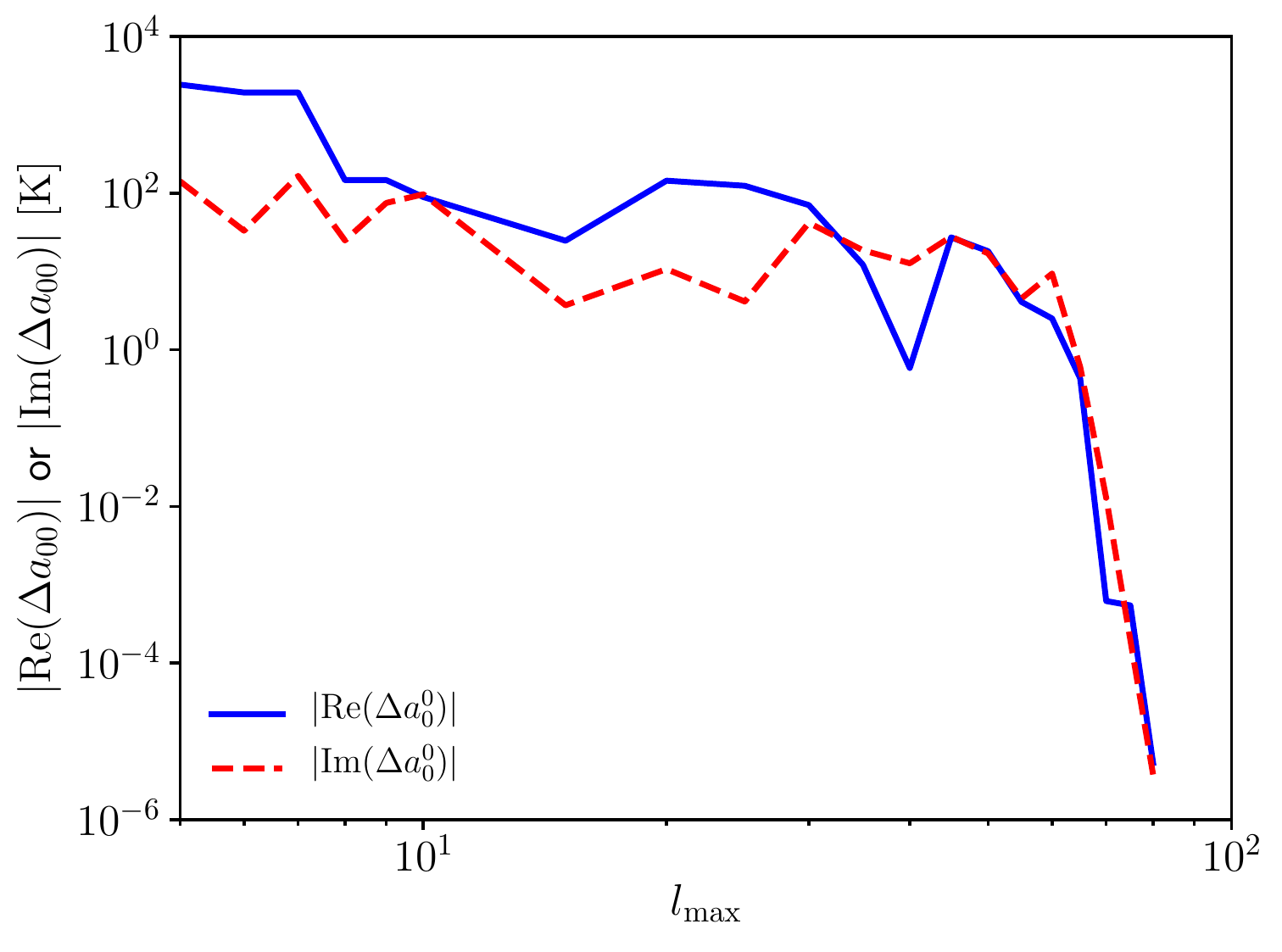}}
\caption{
The absolute value of the difference between $\hat{a}_0^0$ and $a_0^0$, for the real part and the imaginary part respectively, as a function of $l_{\rm max}$.
}
\label{fig:v_vs_l_max}
}
\end{figure}

In Fig. \ref{fig:v_vs_l_max} we plot the error of the solved $\hat{a}_0^0$ (its real and imaginary parts  are the first and the $(N_L+1)$th elements of the vector $\hat{\tilde{\boldsymbol{a}}}$ respectively) compared with the correct $a_0^0$ (note that its imaginary part is zero), as a function of $l_{\rm max}$, for the real part and the imaginary part respectively. We see that the error starts to drop precipitously at $l_{\rm max}\gtrsim 60$. For $l_{\rm max}=80$, the error drops to below $10^{-5}$ K. In summary, a solution of the monopole can be obtained with good accuracy as long as the number of baselines and $l_{\rm max}$ are reasonable.
In the following subsections, we investigate the influence of noise, beam effect, shortest baselines and the baseline distribution on the recovered global spectrum.

\subsection{The noise effect}\label{sec:noise}

Next we simulate the effect of thermal noise by adding a complex Gaussian random  $V_{{\rm N},j}$ on the visibility of the $j$-th baseline. At the low frequency of interest, we assume the system temperature is dominated by the sky temperature. The real and imaginary part of the noise  are independent Gaussian numbers with  mean value zero and 
standard deviation
\begin{equation}
\sigma_V(\nu)=\Omega_B \frac{T_{\rm sky}(\nu)}{\sqrt{2\Delta \nu t_{\rm obs}}},  
\end{equation}
where $T_{\rm sky}(\nu)$ is the mean sky temperature at frequency $\nu$; $\Delta \nu$ is the frequency channel width
and we set it to be 1 MHz throughout this paper; and $t_{\rm obs}$ is the integration time, and the beam coverage 
\begin{equation}
   \Omega_B=\int B_\nu      d \Omega. 
\end{equation}
In this subsection, to clearly show the influence of noise only, we tentatively assume an isotropic beam, so $\Omega_B=4\pi$.

If the instrumental noise is independent of baseline, the final noise level on the recovered global spectrum is (e.g. \citealt{Zhang2016})
\begin{equation}
   \sigma_{\rm N}\sim \frac{1}{\sqrt{4\pi}}\sqrt{ |\tilde{\boldsymbol{Q}}^{-1}
   \boldsymbol{N}_V  (\tilde{\boldsymbol{Q}}^{-1})^\dag|_{00}},
   \label{eq:Delta_T_recovered}
\end{equation}
where $\boldsymbol{N}_V$ is a $2N_L\times 2N_L$ diagonal matrix with 
diagonal elements all equal to $\sigma_V^2$, and ``$^\dag$'' is the Hermitian conjugate (transpose and complex conjugate).

\begin{figure}
\centering{
\subfigure{\includegraphics[width=0.45\textwidth]{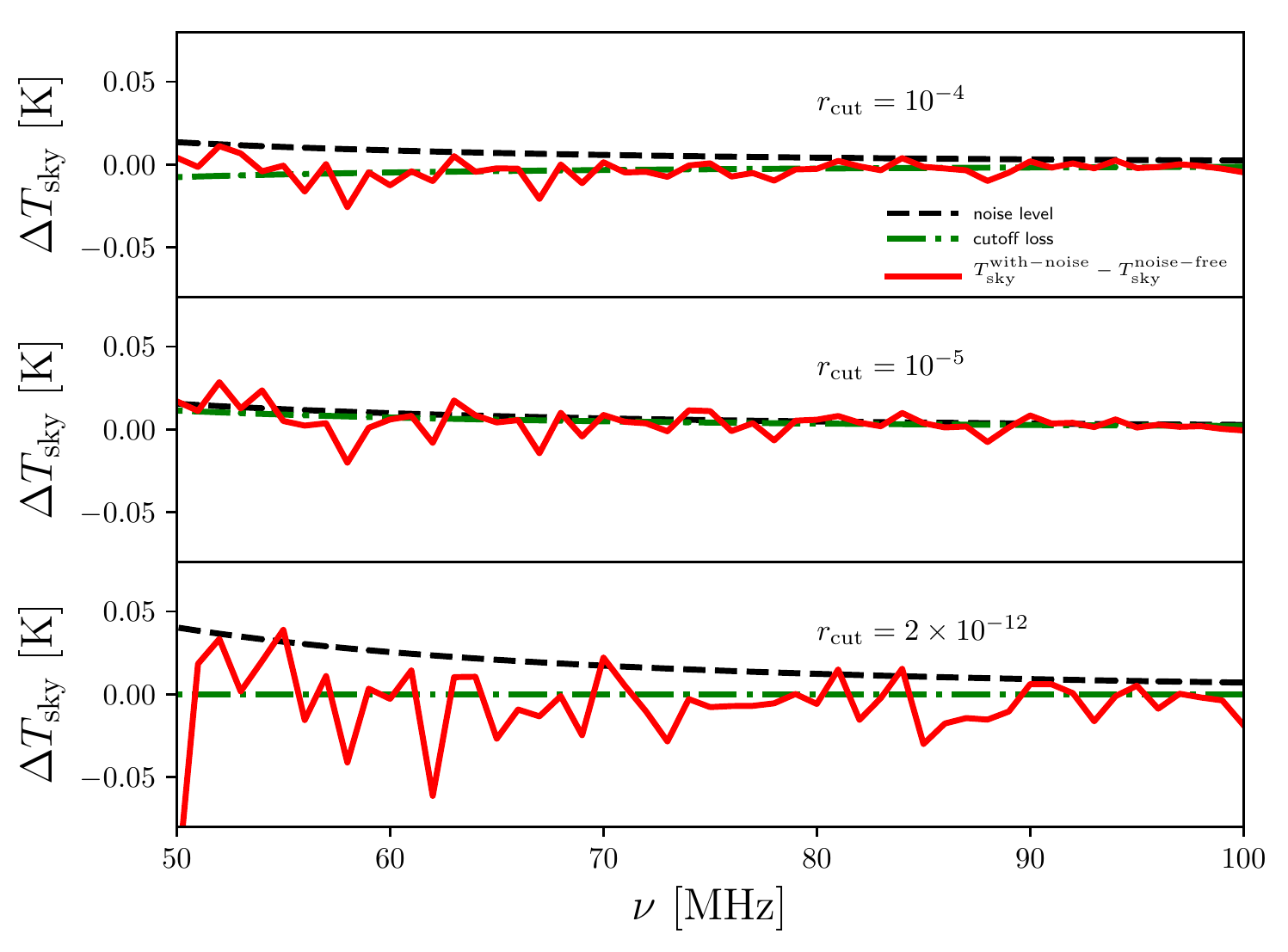}}
\caption{The difference between the recovered sky temperature with and without noise, at different frequencies, for $[b_{\rm min},b_{\rm max}]=[\lambda,10\lambda]$. We adopt $t_{\rm obs}=10^4$ hour. We also plot the noise level estimated from Eq. (\ref{eq:Delta_T_recovered}) and the cutoff loss. From top to bottom, panels correspond to $r_{\rm cut}=10^{-4}$, $10^{-5}$ and $2\times 10^{-12}$ respectively. 
}
\label{fig:T_recovered_random2D_noise}
}
\end{figure}

In the presence of noise, when solving the $\tilde{\boldsymbol{Q}}^{-1}$ we must discard the components corresponding to singular values below a certain cutoff, otherwise the solution would be contaminated by noise (e.g. \citealt{Huang2018}). The contribution to sky temperature below the cutoff is then lost, we shall call this the {\it cutoff loss}.
Cutoff is usually expressed as a fraction $r_{\rm cut}$ of the largest singular value.  According to  Eq. (\ref{eq:Delta_T_recovered}), the noise level on the recovered sky temperature also depends on this cutoff,  and usually the  noise is smaller for larger $r_{\rm cut}$. 

In practice, $r_{\rm cut}$ can only be determined empirically. Fig.  \ref{fig:T_recovered_random2D_noise} shows the difference between the recovered global spectrum with and without noise for $r_{\rm cut}=10^{-4}, 10^{-5}$, $2\times10^{-12}$  respectively. Since the recovered spectrum without noise is almost identical to the input spectrum, this in practice is also the residue  of recovered spectrum in the presence of noise. On the same plot we also show the cutoff loss and the noise level separately. The noise level corresponds to $t_{\rm obs}=10^4$ hour and $[b_{\rm min},b_{\rm max}]= [\lambda,10\lambda]$. From this figure, 
we see that the cutoff loss decreases with decreasing $r_{\rm cut}$, and becomes negligible at $r_{\rm cut}\lesssim10^{-5}$. However the noise level increases with decreasing $r_{\rm cut}$. We find a value  around $\sim10^{-5}$ is roughly applicable to  most of models in this paper. Throughout this paper, we set $r_{\rm cut}=10^{-5}$ unless otherwise specified.

When a suitable $r_{\rm cut}$ is chosen, we expect that as long as the integration time is long enough, the errors are always at low level and are close to the pure instrumental noise. We therefore expect that the 21 cm signal can be extracted  from the recovered global spectrum. Moreover, the conclusion here is not very sensitive to the choice of $r_{\rm cut}$. However, we will see in Sec. \ref{sec:short_b} that for 2D baseline distribution, when $b_{\rm  min}$ is much  loner than the wavelength,  the fluctuations on the recovered sky temperature are huge.
To overcome this, one needs to either use a much longer $t_{\rm obs}$, or use the 3D baselines.

Generally the noise given by Eq. (\ref{eq:Delta_T_recovered}) is much larger than the single antenna noise $ T_{\rm sky}/\sqrt{\Delta \nu t_{\rm obs}}$. This is because it is solved from the equations of visibility for many baselines. When the length of the baseline $b\gg\lambda$, the monopole contribution is much smaller than $b=0$ limit (single antenna). As a result, the derived sky brightness has larger noise. More straightforwardly, the first diagonal element of $\tilde{\boldsymbol{Q}}^{-1} (\tilde{\boldsymbol{Q}}^{-1})^\dag$ is generally much larger than 1. For 4000 random baselines with $\lambda < b <10\lambda$ and $r_{\rm cut}=10^{-5}$, the recovered global sky temperature has noise $\sim15$ ($\sim100$) times larger than the single antenna noise for isotropic beam (Hertz dipole beam).

\subsection{The beam effect}\label{sec:beam}

We then investigate the effect of the beam.  The beam of a Hertz dipole antenna with physical length $L$ 
is \citep{balanis2016antenna},
\begin{equation}
B_\nu( \hat{\boldsymbol{n}})\propto \frac{  [\cos(  \frac{\pi L}{\lambda} \cos\theta' ) -\cos(\frac{\pi L}{\lambda})]^2}{\sin^2\theta'},
\label{eq:dipole_beam}
\end{equation}
where $\theta'$ is the included angle between the direction $\hat{\boldsymbol{n}}$ and the axis along the length of the wire. We assume the antenna is placed horizontally along the W-E direction, without ground blocking. 
We then solve for Eq.~(\ref{eq:V_tilde}), with $Q_{l,j}^m$ given by Eq. (\ref{eq:Q_jlm}) with the above beam function. 
We still adopt baselines in the range $[b_{\rm min},b_{\rm max}]= [\lambda,10\lambda]$, $N_L=4000$ and $l_{\rm max}=80$. 
In visibilities, the fraction of contributions from modes with $l>80$ is generally $\sim10^{-6}-10^{-4}$. In this subsection we neglect the noise.

\begin{figure}
\centering{
\subfigure{\includegraphics[width=0.45\textwidth]{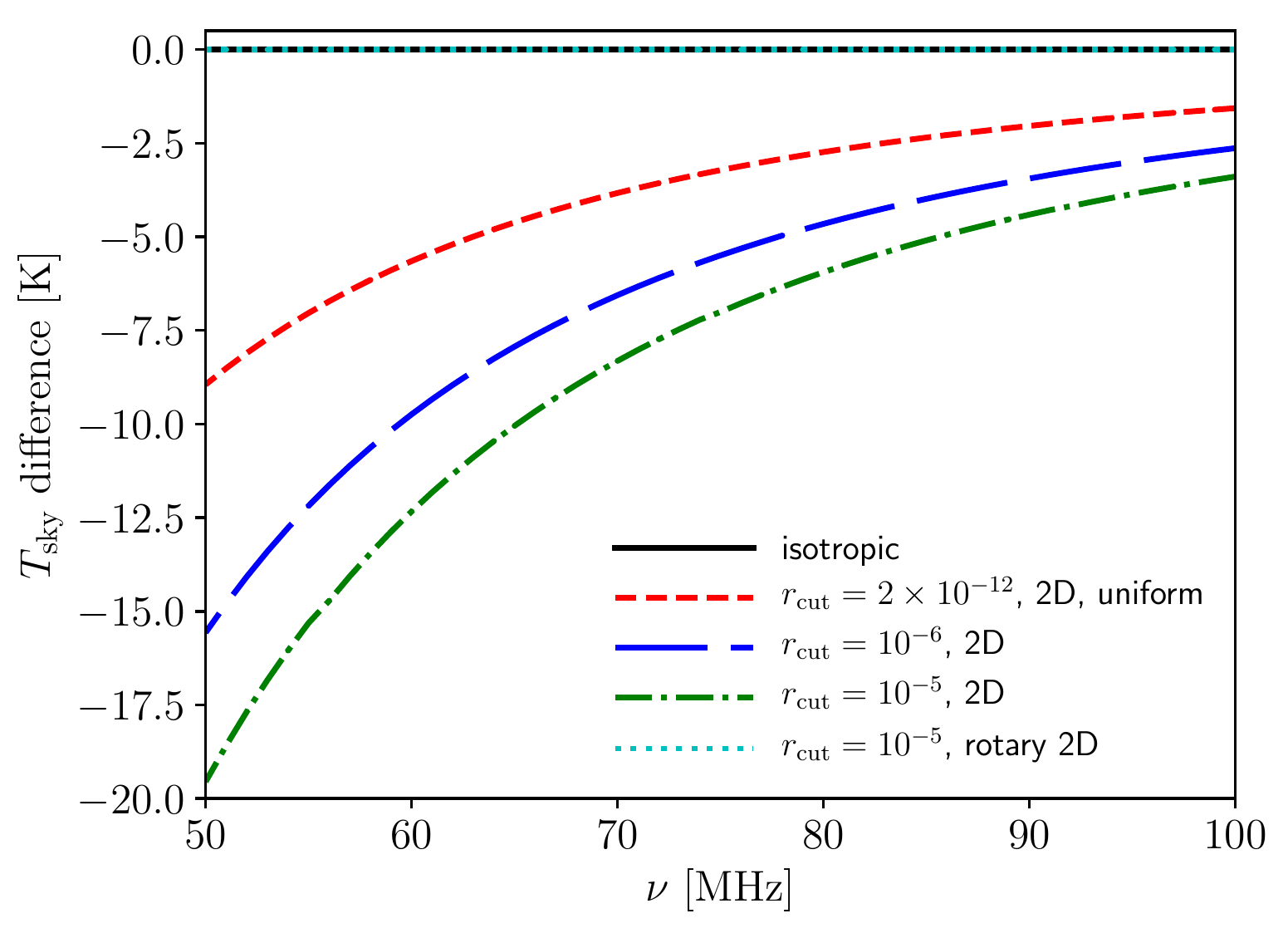}}
\caption{The difference between the recovered global sky temperature for  dipole beam and the input sky model temperature. 2D baselines are randomly distributed in range [$\lambda,10\lambda$]. As comparison we also plot the result for isotropic beam. 
}
\label{fig:T_diff_random2D_beam}
}
\end{figure}

Fig. \ref{fig:T_diff_random2D_beam} shows the difference between the recovered global spectrum and the input sky model spectrum, for the isotropic beam ($B_\nu\equiv 1$) and a dipole beam with $L=1.0$ m respectively (throughout this paper we adopt $L=1.0$ m and all $L<\lambda$ give very similar results). We see that once the dipole beam is introduced, the recovered global spectrum is a bit ($\sim0.1\%$) lower than the actual one, depending on the cutoff. 
The problem is not mitigated by using higher $l_{\rm max}$. Even for a sky with uniform temperature, there is still this underestimation bias. 
Such bias is not produced by the cutoff, because even for the sky with uniform temperature and using a $r_{\rm cut}$ as small as $2\times10^{-12}$, the recovered global spectrum is still biased, 
see the dashed curve in Fig. \ref{fig:T_diff_random2D_beam}.
We also check other forms of beams and find that the results are all biased, although for different beams the bias amplitude is different.    
Generally, a baseline with length $b$ is only sensitive to the modes with $l\lesssim 2\pi b/\lambda$. However, when there is the beam, in the measured visibility the contribution from modes $l\gtrsim 2\pi b/\lambda$ is boosted (the $Q_l^m$s are larger). It is still small compared with the contributions from $l\lesssim 2\pi b/\lambda$, however much larger compared with the isotropic beam. On the other hand, the $\boldsymbol{Q}$ matrix for 2D plane baselines does not contain enough information to constrain these high-$l$ modes. So it is a dilemma: using $l_{\rm max}\sim 2\pi b_{\rm max}/\lambda$ then the ignored contributions from $l>l_{\rm max}$ raise in errors in the solved monopole; using $l_{\rm max} > 2\pi b_{\rm max}/\lambda$ then the high-$l$  modes with  $2\pi b_{\rm max}/\lambda \lesssim l \lesssim l_{\rm max}$ cannot be well determined and they will finally bias the monopole, because to obtain the monopole we solve all $a_l^m$s simultaneously. To support this point, we test that: if we set $l_{\rm max}=2\pi b_{\rm max}/\lambda\approx 60$ and remove the modes with $l > 60$ in the sky model, we can get the unbiased monopole. However if we set $l_{\rm max}=80$ and remove the modes with $l>80$ in the sky model, then the obtained monopole has bias.

This bias would not be a problem as long as it varies smoothly with frequency, which will not be confused with the 21 cm signal when the smooth foreground is subtracted. This is true for fixed baseline distribution, as shown in In Fig. \ref{fig:T_diff_random2D_beam}.
If the baseline distributions differ at different frequencies, it induces extra fluctuations that may confuse the 21 cm signal. In the extreme case, if the global sky temperature at each frequency is recovered from fully independent baseline distributions, then the frequency-fluctuations of the underestimation bias are at the level of $\mathcal{O}(10^{-4})$ of the global sky temperature for $r_{\rm cut}=10^{-5}$. If however we increase the baseline number to $N_L=8000$, they decrease to $\mathcal{O}(10^{-5})$ levels. So by increasing the number of baselines we can reduce the frequency-fluctuations of the bias. Moreover, in Sec. \ref{sec:recover-21cm} we will see that for an array that forms thousands of random baselines such extra fluctuations are well-controlled. It is not a serious problem. For the above reason, in our theoretical investigation we just keep the bias.

As the Earth rotates, the 2D plane of the baselines also rotates. The ultimate method for  obtaining the unbiased solution is to construct visibility equations from all these baseline planes, instead of just using the instantaneous 2D plane. In such case the baselines actually distribute in 3D space. The monopole solution of such equations is unbiased. See the dotted curve in Fig. \ref{fig:T_diff_random2D_beam}. In Sec. \ref{sec:ground2D} we will show the implication of such a method for ground-based 2D antennas.

The Moore-Penrose pseudo-inverse method finds the minimum-norm least squares solution of the equation $\tilde{\boldsymbol{V}}=\tilde{\boldsymbol{Q}} \tilde{\boldsymbol{a}}$, so it minimizes both $ \lVert \tilde{\boldsymbol{V}}-\tilde{\boldsymbol{Q} } \hat{\tilde{\boldsymbol{a}}} \rVert$ and $\lVert \hat{\tilde{\boldsymbol{a}}}\rVert$, where $\lVert \cdot \rVert$ is the  Euclidean norm \citep{zielke1984survey}. In principle this method does not guarantee $\lvert \hat{a}_0^0 \rvert = \lvert a_0^0 \rvert$ (ignore the noise) when there are multi-solutions all have least squares. 
But our calculations find that as long as the baseline coverage is good enough, for many beam forms, 
the solved $\lvert \hat{a}_0^0 \rvert$ is indeed a good estimator of $\lvert a_0^0 \rvert$ if ignore numerical errors. Examples are given  in Sec.  \ref{sec:beam_dependence}.

\subsection{Noise \& beam}\label{sec:noise_and_beam}

\begin{figure}
\centering{
\subfigure{\includegraphics[width=0.45\textwidth]{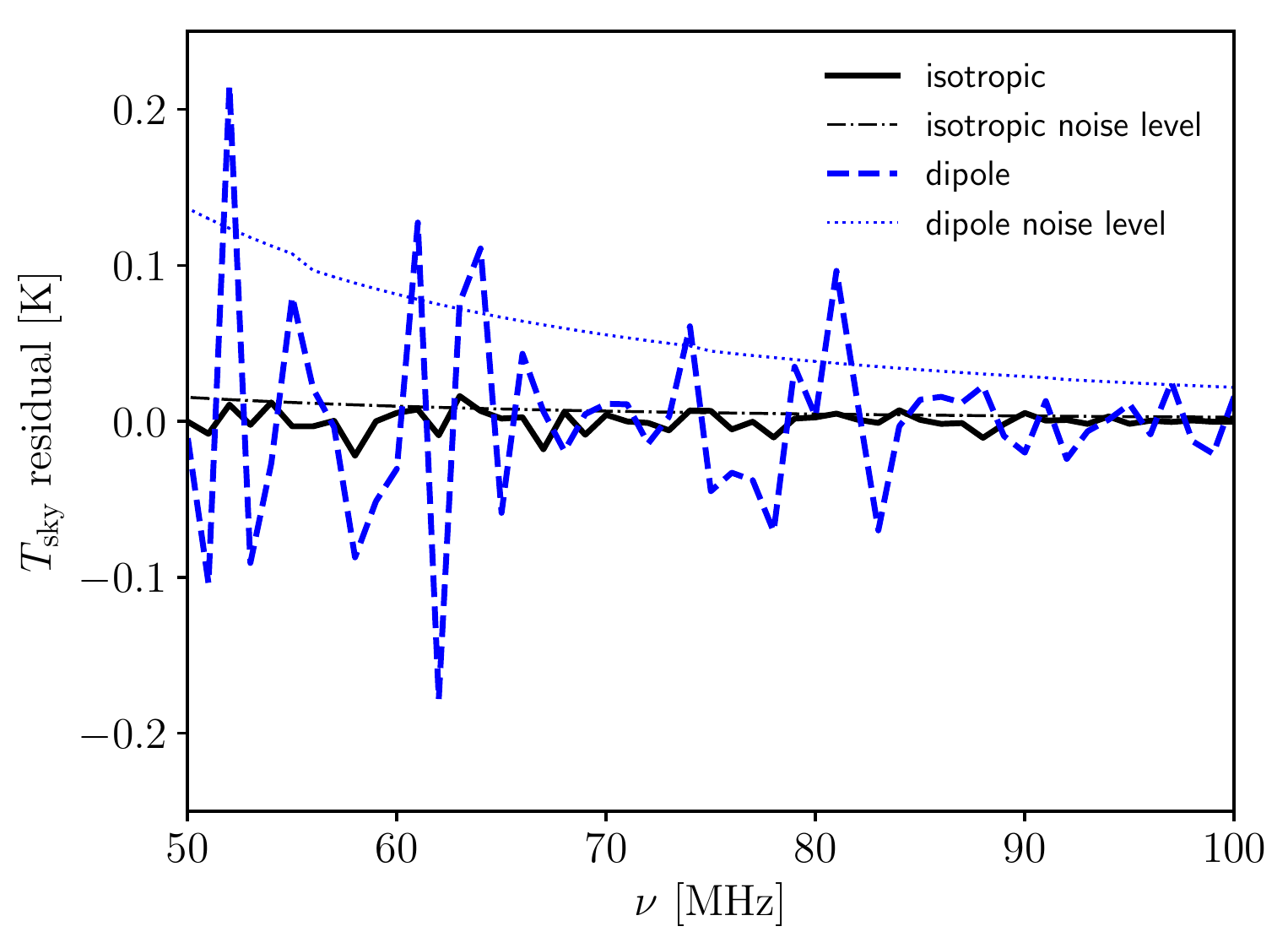}}
\caption{The residuals after removing the best-fit smooth component from the recovered global spectrum, for 2D random baselines with different beams as labeled. Baselines are in range [$\lambda,10\lambda$] and the noise levels correspond to  $t_{\rm obs}=10^4$ hour. Here we adopt $r_{\rm cut}=10^{-5}$. As reference we also plot the noise level by thin lines.
}
\label{fig:T_residuals_random2D_noise_beam}
}
\end{figure}

We then check the smoothness of the recovered  global spectrum in the presence of both beam and noise effect. The smoothed component   (foreground) is  described  as \citep{GSM_2008}
\begin{equation}
T_{\rm FG}(\nu)=T_0\left( \frac{\nu}{\nu_0}    \right)^{\beta(\nu)},
\label{eq:T_FG}
\end{equation}
where the spectrum index is a polynomial
\begin{equation}
\beta(\nu)=\sum_{i=0}^{N_{a}} a_i [\ln (\nu/\nu_0)]^i.
\end{equation}
We adopt $N_a=5$. In Fig.  \ref{fig:T_residuals_random2D_noise_beam} we show the residuals after subtracting the best-fit smooth component from the recovered global spectrum, for the isotropic beam and dipole beam. 
All residual fluctuations are well controlled (dominated by pure instrumental noise and can be reduced by simply increasing integration time).

\subsection{3D baselines}\label{sec:3Dbaselines}

So far we have investigated  the measurement taken with 2D planar baselines. The baselines can also be non-planar, i.e. distributed in  the 3D  space, and for the 21 cm global spectrum extraction this may be important. 
First, we find that when the beam is isotropic, for 3D baselines the cutoff loss of the recovered global sky temperature is much smaller than 2D, for the same $r_{\rm cut}$. For example, when $r_{\rm cut}\lesssim10^{-5}$, the cutoff loss for both 2D and 3D baselines is negligible.
However, when $r_{\rm cut}$ is as high as $10^{-1}$, the cutoff loss is $\sim40\%$ for 2D baselines, while just $\sim0.1\%$ for 3D baselines. So the 3D baseline distribution has an obvious advantage. It has high immunity to noise since it allows to choose a larger $r_{\rm cut}$.
This is because for the 2D baseline distribution, the singular values of the $\tilde{\boldsymbol{Q}}$ matrix drop dramatically above the ordinal number $\sim1000$, while for the 3D baseline distribution they just decrease gently until $\sim4000$. So for the same criterion $r_{\rm cut}$ the 2D baseline distribution loses more principle components.
More intuitively, since we solve all spherical harmonic coefficients simultaneously, it is necessary to have independent visibility measurements from baselines with different lengths and orientations.
The 3D baseline distribution has one more orientation freedom compared with the 2D planar distribution, allowing it to provide more distinctive visibility measurements.

\begin{figure}
\centering{
\subfigure{\includegraphics[width=0.45\textwidth]{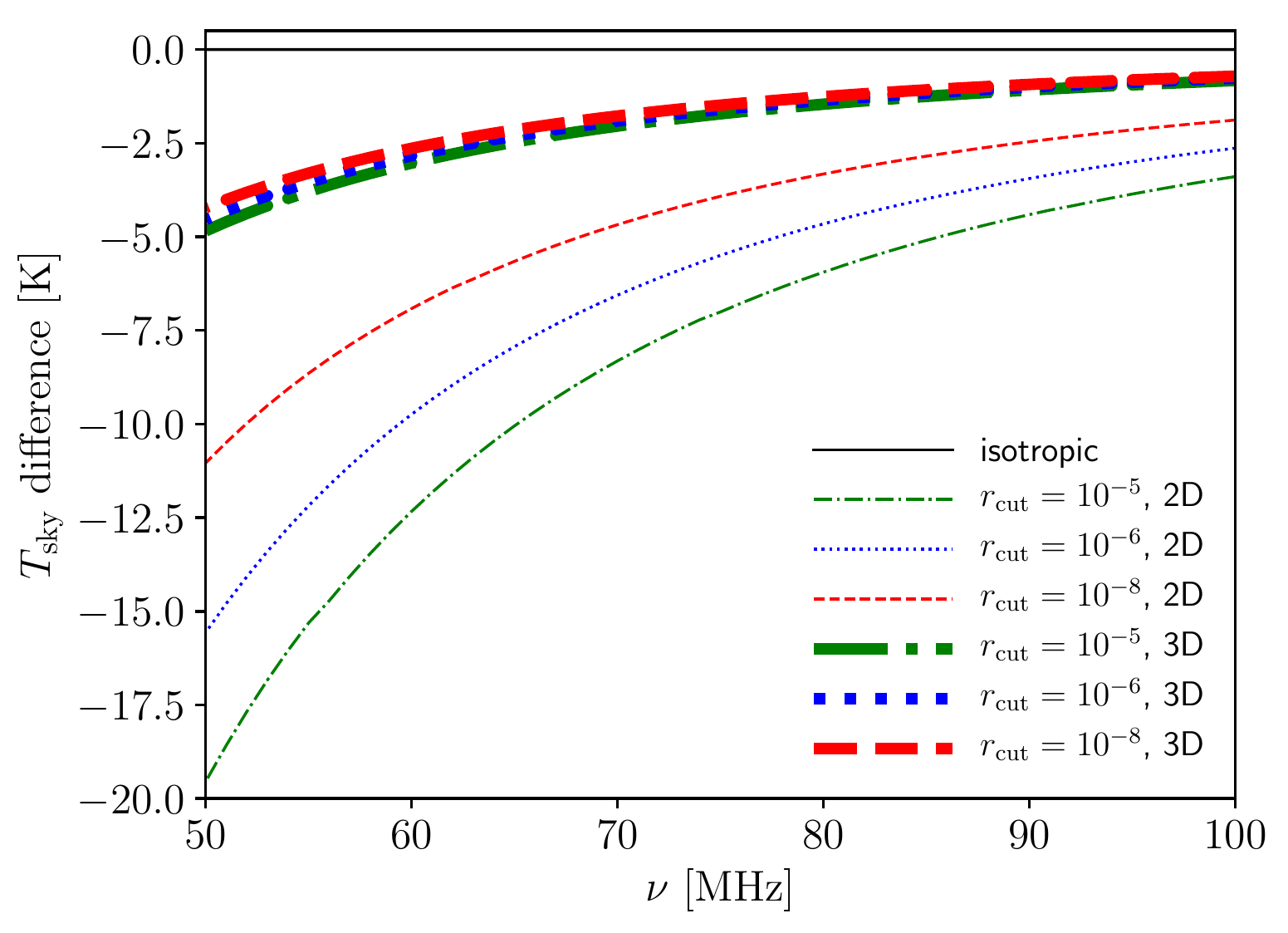}}
\caption{The difference between the recovered sky temperature and the input sky temperature, for 2D (thin lines) and 3D (thick lines) baselines. We take $[b_{\rm min},b_{\rm max}]=[\lambda,10\lambda]$.  
}
\label{fig:T_diff_random3D}
}
\end{figure}

We then consider the dipole beam.
In Fig. \ref{fig:T_diff_random3D}
we show  the difference between the recovered global spectrum and the input sky model temperature, for 3D baseline distribution.  As comparison we also plot the results for the 2D baselines. Both 3D and 2D baseline distributions have the same $[b_{\rm min},b_{\rm max}]=[\lambda,10\lambda]$, 
same number of baselines $N_L=4000$, same dipole antennas.  We see that for the same beams and the same number of baselines, the recovered global spectrum in 3D baselines has smaller underestimation bias than the input sky temperature. 
The bias depends on the included angle between the antenna wire and the baseline plane. When the wire is vertical to the baseline plane (the included angle is $\pi/2$, the minimum response direction of the beam is vertical to the baseline plane), the bias is smaller than when the wire is parallel to the baseline plane (the included angle is 0, the minimum response direction of the beam is parallel to the baseline plane). 
The 3D baseline distribution has many baseline planes with included angles between $0-\pi/2$, as a result, the bias is smaller than when the included angle is 0 for 2D baselines.

\subsection{The shortest baselines}\label{sec:short_b}

\begin{figure}
\centering{
\subfigure{\includegraphics[width=0.45\textwidth]{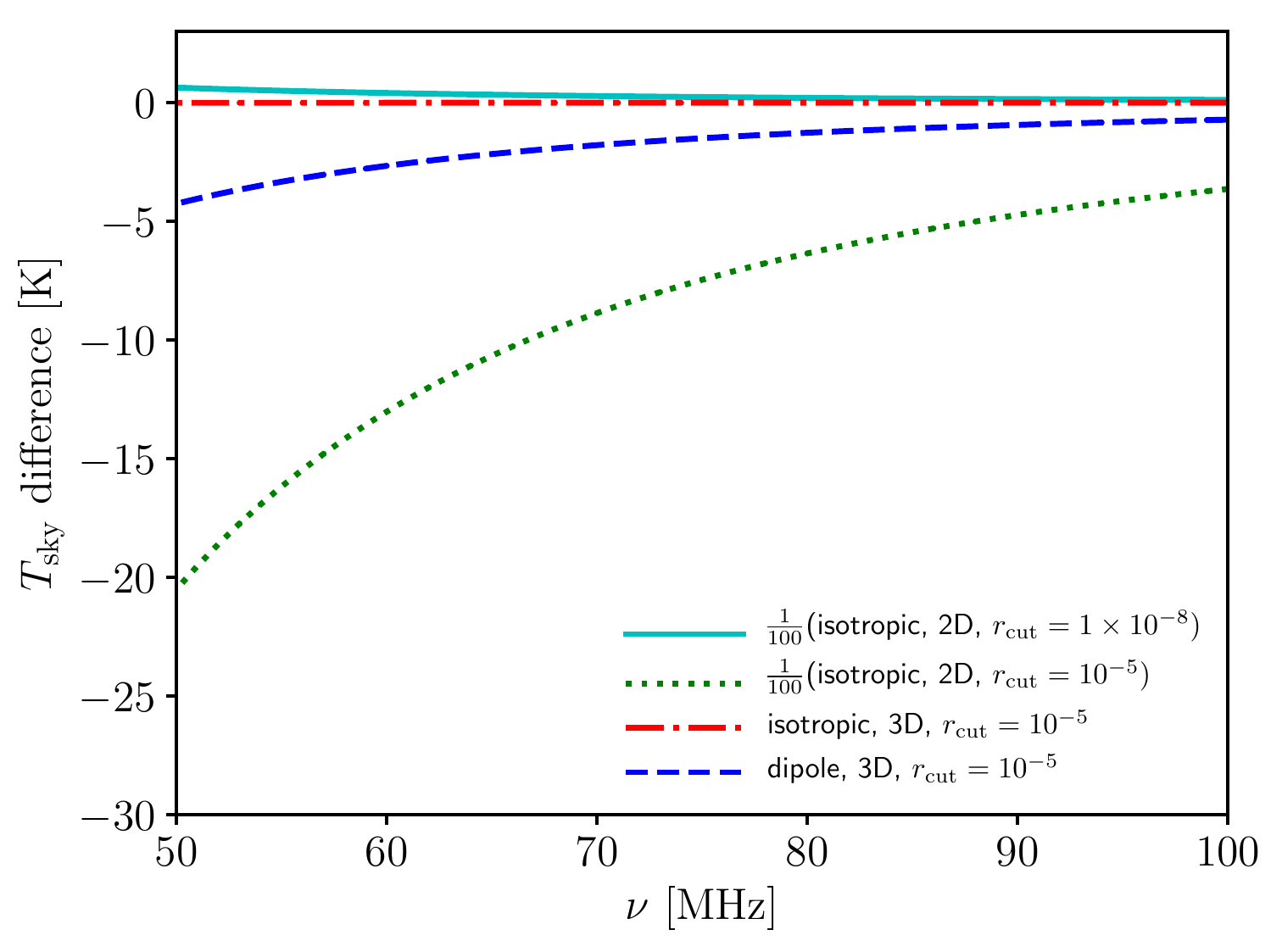}}
\caption{
The difference between the recovered global sky temperature and the input sky model temperature for models with $[b_{\rm min},b_{\rm max}]=[3\lambda,10\lambda]$. For the 2D isotropic beam we multiply a factor $\frac{1}{100}$, so that all curves can be displayed clearly in same panel.    
}
\label{fig:T_diff_shortest_baselines}
}
\end{figure}

In all above calculations the shortest baselines are equal to wavelength. Can we still recover the sky temperature if the shortest baselines are longer?
This is an important question because in practice, the spacing between the antennas of interferometer are limited by always the physical size of the antenna, usually this spacing is larger than a wavelength, and more over, antennas nearby could couple each other.

We find that when $[b_{\rm min},b_{\rm max}]=[3\lambda,10\lambda]$, for the 2D baselines the recovered global spectrum has large error even when the beam is isotropic, no matter how to choose the $r_{\rm cut}$, see the solid and dotted curves in Fig. \ref{fig:T_diff_shortest_baselines}.
Our spherical harmonic coefficients are solved simultaneously from visibilities measured by baselines with various lengths.
Although long baselines contain the information of both low-$l$ and high-$l$ modes, short baselines are more important because they are only sensitive to the low-$l$ modes, therefore can help to anchor the $\hat{a}_0^0$.
In the absence of short baselines, the solution $\hat{a}_0^0$ relies on solutions of high-$l$ modes, therefore is sensitive to the choice of $r_{\rm cut}$ that is a criterion for discarding higher order principle components in $\tilde{\boldsymbol{Q}}$.

On the other hand, for an array with 3D spatial distribution of baselines, the performance is good. The recovered global spectrum is not influenced even if the minimum baselines increase from $\lambda$ to $3\lambda$ for both isotropic beam and dipole beam. 
Because 3D baselines contain more valid independent visibility measurements, see Sec. \ref{sec:3Dbaselines}.
This is another reason we  strongly  recommend to use 3D baseline distributions.

Moreover, we also check the noise level. For the 2D baselines, if the minimum baselines increase from $\lambda$ to $3\lambda$, the noise level increases by more than two orders of magnitude. Even if the bias can be corrected, such high noise level makes it difficult to extract 21 cm signal from the recovered global spectrum in this model. Fortunately, for the 3D baselines, the noise level is quite similar when $b_{\rm min}=\lambda$ and $b_{\rm min}=3\lambda$.

 Regarding the longest baselines, we have always taken $b_{\rm max}=10\lambda$ in such simulations. In the presence of short baselines, longer baselines are not necessary because their response to global spectrum is negligible \citep{Presley2015}. Throughout this paper, when $b_{\rm min}= \lambda$, all conclusions are similar if we replace $b_{\rm max}=10\lambda$ with $b_{\rm max}=2-3\lambda$. however using $b_{\rm max}=10\lambda$ slightly improve the bias and noise. Moreover, if for some reason one has to remove baselines $\lesssim3\lambda$, for example to suppress the cross-coupling, one can still recover the global spectrum from visibilities observed by baselines with $3\lambda < b <10\lambda$ for 3D baseline distribution.

\section{Array configuration and 21 cm signal recovery}\label{sec:recover-21cm}

In this section, we investigate the feasibility of recovering the global spectrum from visibilities  measured by interferometer with realistic array configurations, and the feasibility of  extracting  21 cm signal  from the  recovered global spectrum.

\subsection{The 21 cm signal from cosmic dawn}\label{sec:21cm-signal}

Here we generate mock 21 cm signal maps that will be used in following simulations.
At cosmic dawn, the reionization is still negligible, but  the X-ray heating and Ly$\alpha$ coupling can be highly inhomogeneous. It produces an inhomogeneous spin temperature field and changes the 21 cm signal field significantly (e.g. \citealt{Ghara2015MNRAS}).  
In this paper, however, to avoid introducing more model uncertainties and for simplicity, we assume 
the 21 cm signal traces the neutral Hydrogen, and neutral Hydrogen traces the dark matter. 
As a test signal this is acceptable.
The 21 cm power spectrum 
\begin{align}
P_{21}(k,z)&=b_{21}^2(z)P_{\rm H}(k,z)   \nonumber \\
&=b_{21}^2(z)P_{\rm m}(k,z),
\end{align}
where $P_{\rm H}$ and $P_{\rm m}$ are Hydrogen and dark matter power spectra respectively. The bias
\begin{equation}
b_{21}(z)=1.18\left(\frac{h}{0.7}\right)\left(\frac{n_{\rm H}}{10^{-7}}\right) \left[\frac{(1+z)^2}{E(z)}\right]~~~[\rm mK],
\end{equation}
where $n_{\rm H}$ is the cosmic Hydrogen density in the present Universe, in units of cm$^{-3}$.

The 21 cm angular power spectrum is (e.g.  \citealt{Loeb2004PRL})
\begin{equation}
C_l(z)=\frac{2}{\pi}\int dk k^2P_{21}(k,z) \mathcal{J}^2_l(kr(z)).     
\label{eq:21_Cl}
\end{equation}

Our purpose is to test whether the 21 cm global spectrum can be recovered from sky temperature measured with interferometer array.  Therefore we approximate the 21 cm global spectrum at cosmic dawn by a Gaussian form, since it is easy for parameterization.
\begin{equation}
\delta T_{21}(\nu)=A\exp\left[-\frac{(\nu-\nu_{21})^2}{2\sigma_{21}^2}\right], 
\label{eq:21_global}
\end{equation}
where $A$, $\nu_{21}$ and $\sigma_{21}$ are three free parameters. Throughout this paper, we take $A=-0.5$ K, $\nu_{21}=75$ MHz and $\sigma_{21}=5$ MHz as our input test 21 cm signal.

We make the fluctuations map for 21 cm signal as random realization the angular power spectrum Eq. (\ref{eq:21_Cl}) and then add the global signal from Eq. (\ref{eq:21_global}). This  is the mock 21 cm signal map that will be added to the sky model.

To extract the 21 cm signal, one  fits  the  recovered  global spectrum with  foreground  plus 21 cm signal 
simultaneously.
The parameters of foreground and 21 cm signal are obtained by performing Markov Chain Monte Carlo (MCMC) analysis that minimizing
\begin{equation}
\chi^2 =\sum_i \frac{[T_{\text{FG}}(\nu_i)+T_{21}(\nu_i)-\hat{T}_{\text{sky}}(\nu_i)]^2}{\sigma^2_{\nu_i}},
\end{equation}
where $\hat{T}_{\rm sky}(\nu_i)$ is the recovered global sky temperature at frequency $\nu_i$. The fitting is accomplished by using the EMCEE3 Python package \citep{Foreman-Mackey2013PublAstronSocPac}.

\subsection{A ground-based  2D telescope array}\label{sec:ground2D}

We first consider a traditional ground based planar (2D) array. The array is composed of 
$N_{\rm antenna}=400$ dipoles
randomly distributed inside a circle with radius $r_{\rm array}=60$ m. However the physical distance between any neighboring antennas is required to be $>2$ m. This is to maximize the randomness of baseline distribution and $u$-$v$ coverage. {\it At each frequency we only choose the baselines with $\lambda < b<10\lambda$} for use in the equation. Here we assume the ground is opaque and blocks half of the sky under the horizon, so that only the part of sky above horizon will contribute, and we ignore the reflection by the ground. Alternatively, the ground can be made to be totally reflective at this frequency. For carefully designed and mounted antennas,  the reflection will not produce new side lobe in the antenna beam \citep{straw2007arrl}.

The antennas are assumed to be fixed.  As the Earth rotates, the beam moves gradually along the celestial parallel. So at different time snapshot, the array measures different hemisphere. For the simulation, we take 24 snapshots each day and derive the global spectrum at each snapshot. The  average of the 24 global spectrum snapshots is used to represent the all-sky global spectrum. The noise level at each snapshot is Eq. (\ref{eq:Delta_T_recovered}). The final noise level  is actually  $1/\sqrt{24}$ of the mean noise of all the snapshots.

\begin{figure}
\centering{
\subfigure{\includegraphics[width=0.4\textwidth]{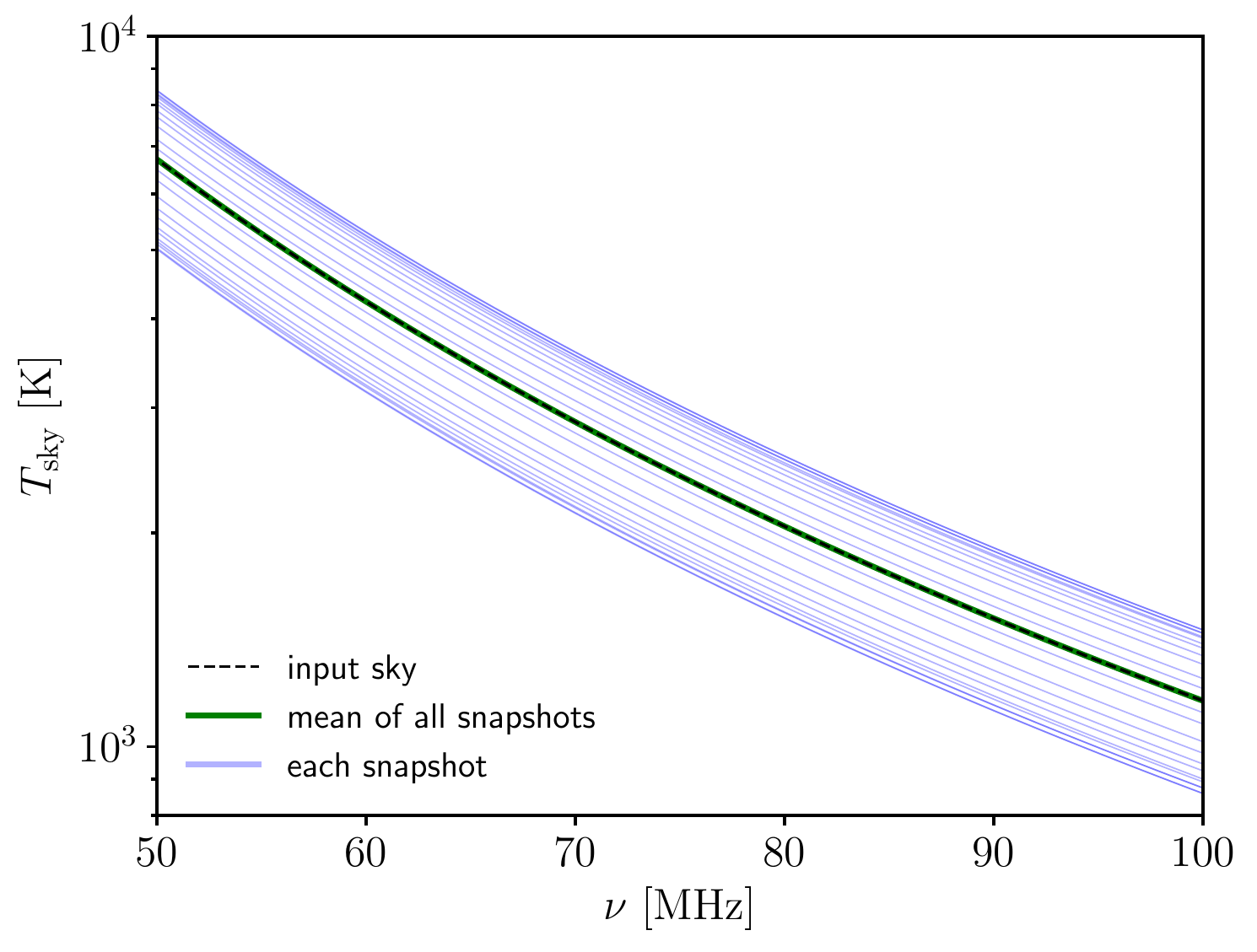}}
\subfigure{\includegraphics[width=0.4\textwidth]{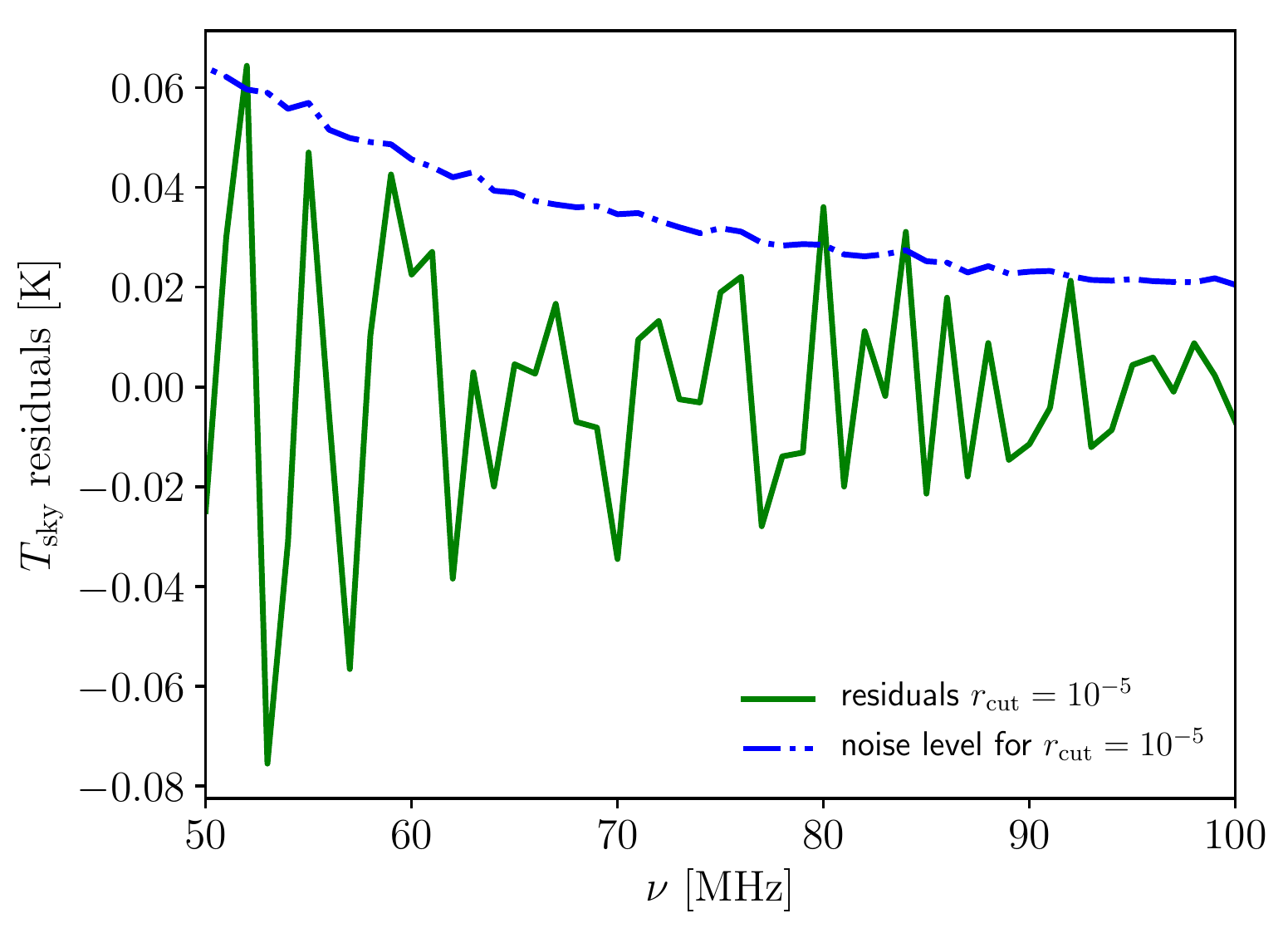}}
 \caption{
{\it Top:} The recovered global sky temperature for a ground-based 2D dipoles array. Curves with light blue color are the global spectrum at each time  snapshot. The curve   with  dark green color is the mean of these. As comparison we  plot  the input sky temperature of the sky model. {\it Bottom:} The residuals after removing the best-fit smooth component from the recovered global spectrum (without noise and 21 cm signal), for $r_{\rm cut}=10^{-5}$. As comparison we also plot the noise level. 
}
\label{fig:T_recovered_ground2D}
}
\end{figure}

In Fig. \ref{fig:T_recovered_ground2D} we plot the recovered sky temperature as a function of frequency at the 24 snapshots and their average. Each snapshot 
accomplishes $10^2$ hour 
integration time, and we adopt $r_{\rm cut}=10^{-5}$. To correct for the block effect all recovered temperature is multiplied by a factor $1/f_{\rm unblocked}$, where $f_{\rm unblocked}$ is the fraction of the unblocked sky area, and $f_{\rm unblocked}$ simply equals $1/2$. Although there is a spread of recovered  temperature values for each snapshot, their mean is quite close to the global spectrum value of the input.

As a first check, we examine the fluctuations in the recovered global spectrum when noise and 21 cm signal are not included. We fit the mean  global spectrum  by Eq. (\ref{eq:T_FG}) with $N_{a}=5$. The residuals after removing the best-fit curve from the recovered global spectrum are shown in the bottom panel of  Fig. \ref{fig:T_recovered_ground2D}. We see that there are still  fluctuations in residuals.

The residual fluctuations are caused by this:  When the baselines have fixed physical length, at different wavelength 
the number and distribution of valid baselines are  different,
because we only use the baselines $\lambda<b<10\lambda$. As a result,  at different wavelength the recovered temperature could slightly deviate from the idea random model as in Sec. \ref{sec:algorithm}. Using the cutoff $r_{\rm cut}$ makes it worse. Because at different wavelength $\tilde{\boldsymbol{Q}}$  has different singular values. The  $\tilde{\boldsymbol{Q}}^{-1}$ derived by discarding the singular values below $r_{\rm cut}$ at different wavelength results in signal loss, as we noted in Sec. \ref{sec:noise}. 
The cutoff loss also changes  as the Earth rotates. If  such loss is just a  smooth function of frequency,  then it is not a problem for extracting the 21 cm signal. Unfortunately,  the cutoff loss has fluctuations, and it depends on not only the array configuration but also on the temperature distribution on the sky,  so for each snapshot the loss is different.    
Whether such residual fluctuations are problem depends on the baseline coverage. From the bottom panel of Fig. \ref{fig:T_recovered_ground2D}, for our adopted array parameters the fluctuations in residuals are smaller than the noise. It is not a problem here.
The fluctuations can be further reduced if the baseline coverage is more close to the idea random distribution.

As checkup, we also make a simulation for a more compact array: 200 antennas are distributed inside a circle with a radius $20$ m, and at each frequency we only pick up the baselines with $\lambda < b < 3\lambda$. For this array we get quite similar noise level and residual fluctuations of the recovered global spectrum.

\begin{figure}
\centering{
\subfigure{\includegraphics[width=0.45\textwidth]{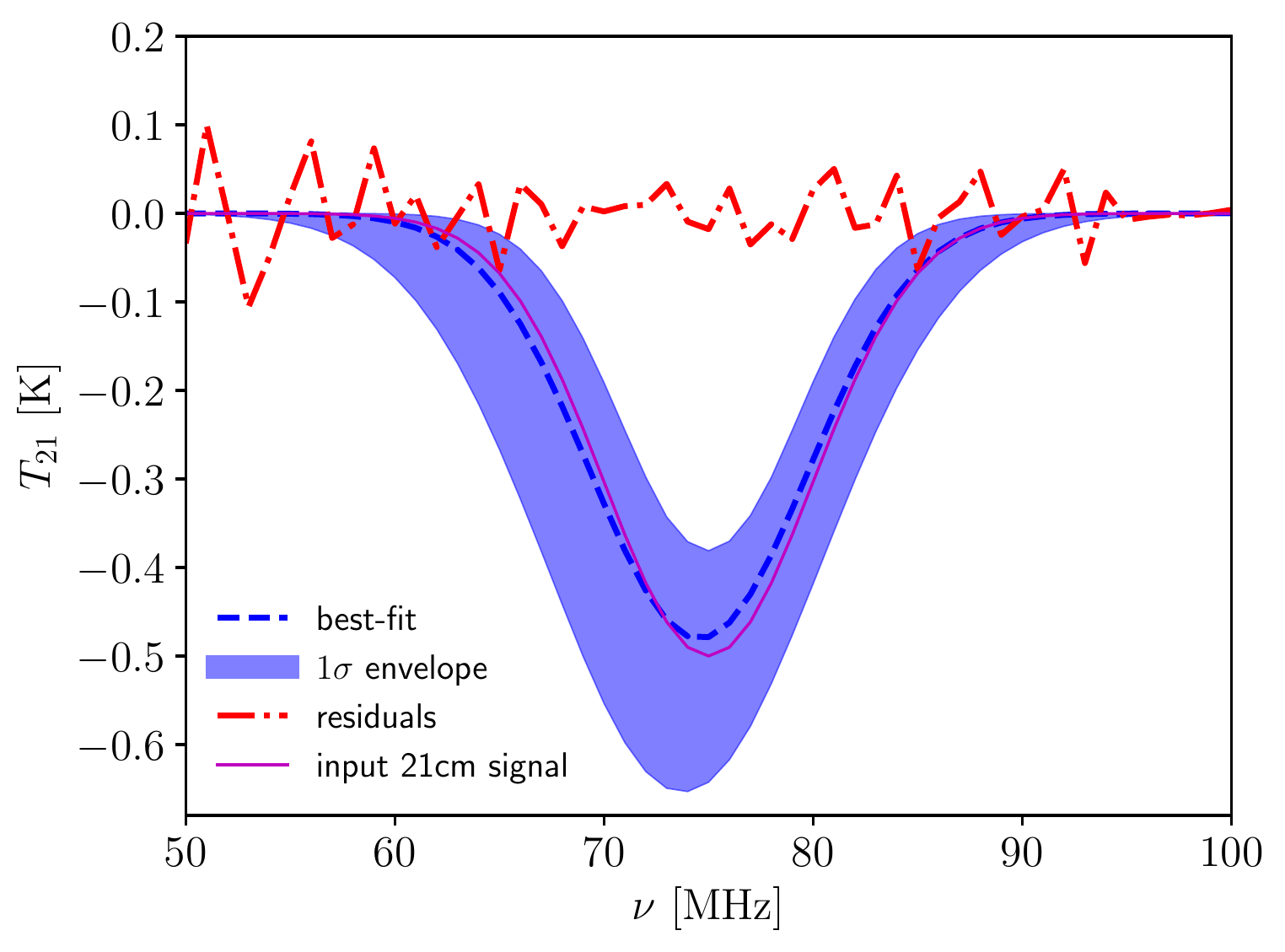}}
\caption{The best-fit 21 cm signal for the recovered global spectrum by 2D ground based array (blue dashed), and the residuals after we remove the best-fit foreground and 21 cm signal from the recovered global spectrum (red dashed-dotted). We also plot the envelope of all 21 cm curves that fit the data with uncertainties $<1\sigma$.
For comparison we also plot the input 21 cm signal (thin black magenta line).  
}
\label{fig:T_21cm_ground2D}
}
\end{figure}

We next add the 21 cm signal to the input sky  model,  recover the global spectrum  again  and fit the foreground plus 21 cm signal of Eq. (\ref{eq:21_global}) simultaneously. To reduce the degeneracy between the foreground and the 21 cm signal, here we use the form in \citet{Shi2022_global} for the foreground, 
\begin{equation}
T_{\rm FG}(\nu)=\left(\frac{\nu}{\nu_0}\right)^{-2.5}\left[T_0+\sum_{i=1}^{4}a_i\left( \log \frac{\nu}{\nu_0} \right)^i \right].    
\end{equation}
They have proved that this formula performs better than the traditional polynomial form like Eq. (\ref{eq:T_FG}). More details could be found in \citet{Shi2022_global}. We use the prior: $-2.0$ K$ <A<$0.0 K, 50 MHz $<\nu_{21}<$ 100 MHz and 1 MHz $<\sigma_{21}<$ 20 MHz. In  Fig. \ref{fig:T_21cm_ground2D} we plot the best-fit 21 cm curve, the residuals after we remove the best-fit foreground and 21 cm signal from the recovered global spectrum, and the envelope of all 21 cm signal that fits the recovered global spectrum within $1\sigma$ deviation. The marginalized parameters for the 21 cm signal are:   
$A=-0.49_{-0.05}^{+0.04}$ K, $\nu_{21}=74.6_{-0.3}^{+0.3}$ MHz and $\sigma_{21}=5.3_{-0.4}^{+0.4}$ MHz, agree with the input 21cm signal very well.

\begin{figure}
\centering{
\subfigure{\includegraphics[width=0.45\textwidth]{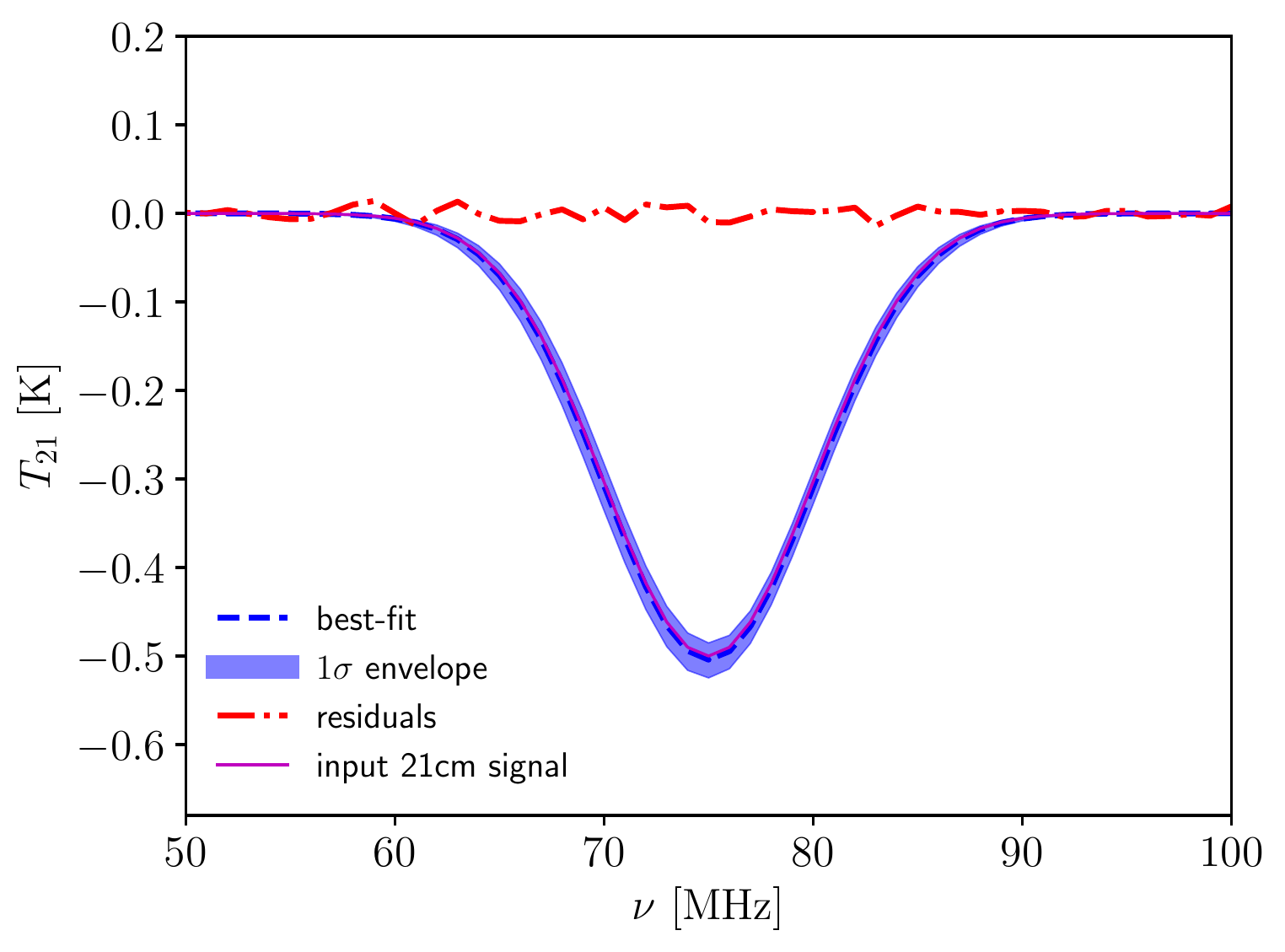}}
\caption{Similar to Fig. \ref{fig:T_21cm_ground2D} except that in this figure the 21 cm global spectrum is recovered from the visibility equations of baselines of different snapshots. So the baselines are actually have 3D distribution.
}
\label{fig:T_21cm_ground2D_Earth_rotation}
}
\end{figure}

Instead of recovering the global spectrum from just instantaneous baselines, we can also put the baselines of different snapshots together to build the visibility equations. As pointed out in Sec. \ref{sec:beam} in this case the baselines are actually in 3D space and the bias problem is solved. In Fig. \ref{fig:T_21cm_ground2D_Earth_rotation} we show the recovered 21 cm global spectrum in such case. Since the visibility equations contain baselines of different snapshots, we can use fewer antennas and snapshots. For Fig. \ref{fig:T_21cm_ground2D_Earth_rotation} we only use 100 antennas and 8 snapshots. 
The recomved 21 cm signal is 
$A=-0.505_{-0.007}^{+0.007}$ K, $\nu_{21}=75.01_{-0.05}^{+0.05}$ MHz and $\sigma_{21}=5.05_{-0.07}^{+0.07}$ MHz.
Obviously, the recovered 21 cm global spectrum is better than using only instantaneous baselines in Fig. \ref{fig:T_21cm_ground2D}.

\subsection{A ground-based 3D telescope array}\label{sec:ground3D}

\begin{figure}
\centering{
\subfigure{\includegraphics[width=0.45\textwidth]{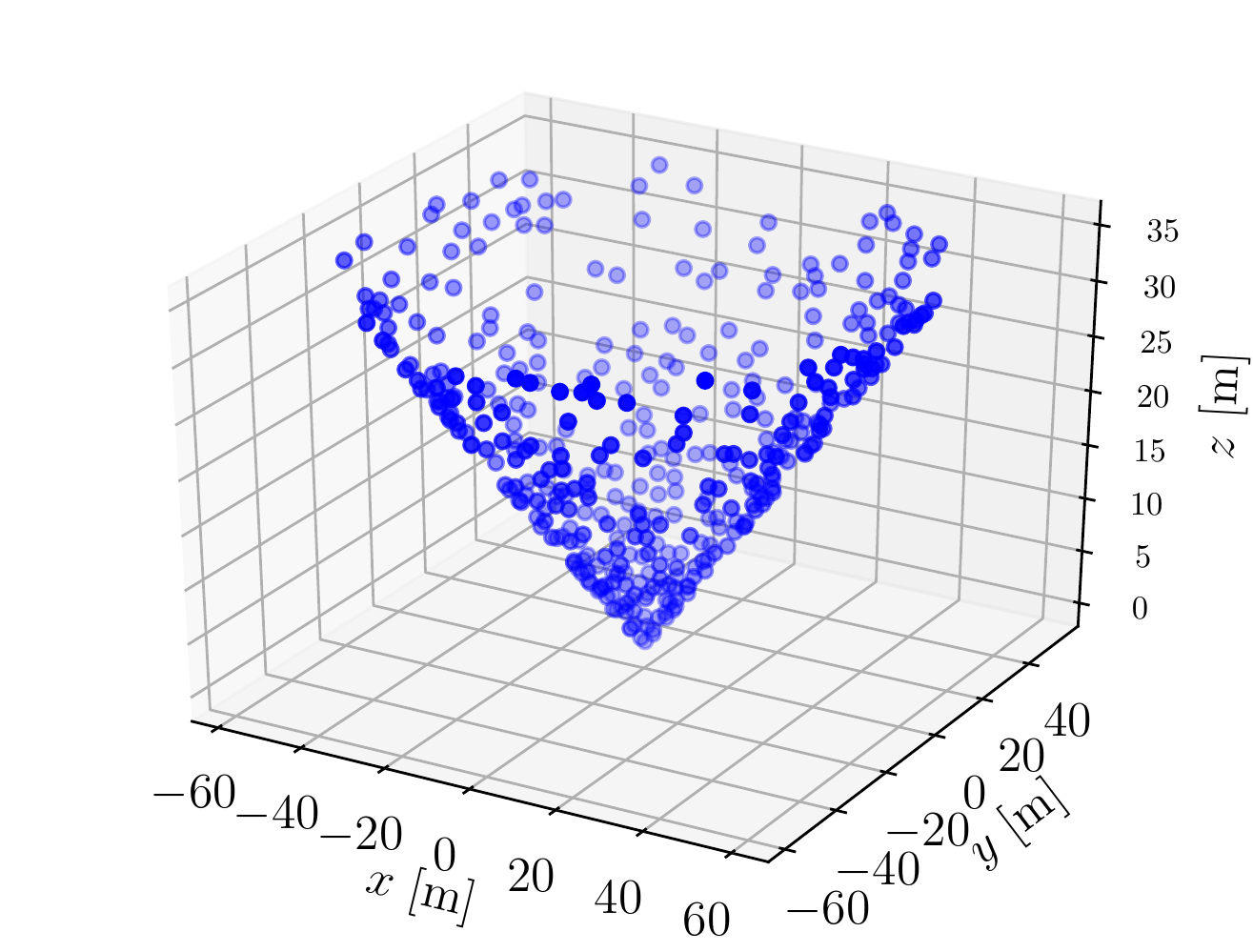}}
\caption{The location of each antenna in the ground-based 3D interferometer array.  
}
\label{fig:xyz_antenna}
}
\end{figure}

We then consider a ``3D'' telescope array on the Earth, in a deep conical well. The wall of the well has inclination angle $\alpha_{\rm wall}=30^\circ$ with   respect to the horizon.  Similar to Sec. \ref{sec:ground2D}, all 400 antennas are randomly distributed inside a circle with radius $60$ m. However each antenna also has a height $r\tan\alpha_{\rm wall}$, where $r$ is horizontal distance to the center of the well. The physical distance between any neighbouring antennas must be $>$2 m. The array  configuration is shown in Fig. \ref{fig:xyz_antenna}. Again at each frequency we only choose the baselines with $\lambda < b<10\lambda$. In computing the visibility  we  block the sky regions with altitude $<\alpha_{\rm wall}$. For this array $f_{\rm unblocked}=0.25$. Same to Sec. \ref{sec:ground2D}, we take 24 snapshots each day, and each snapshot has $10^2$ hour integration time.

\begin{figure}
\centering{
\subfigure{\includegraphics[width=0.4\textwidth]{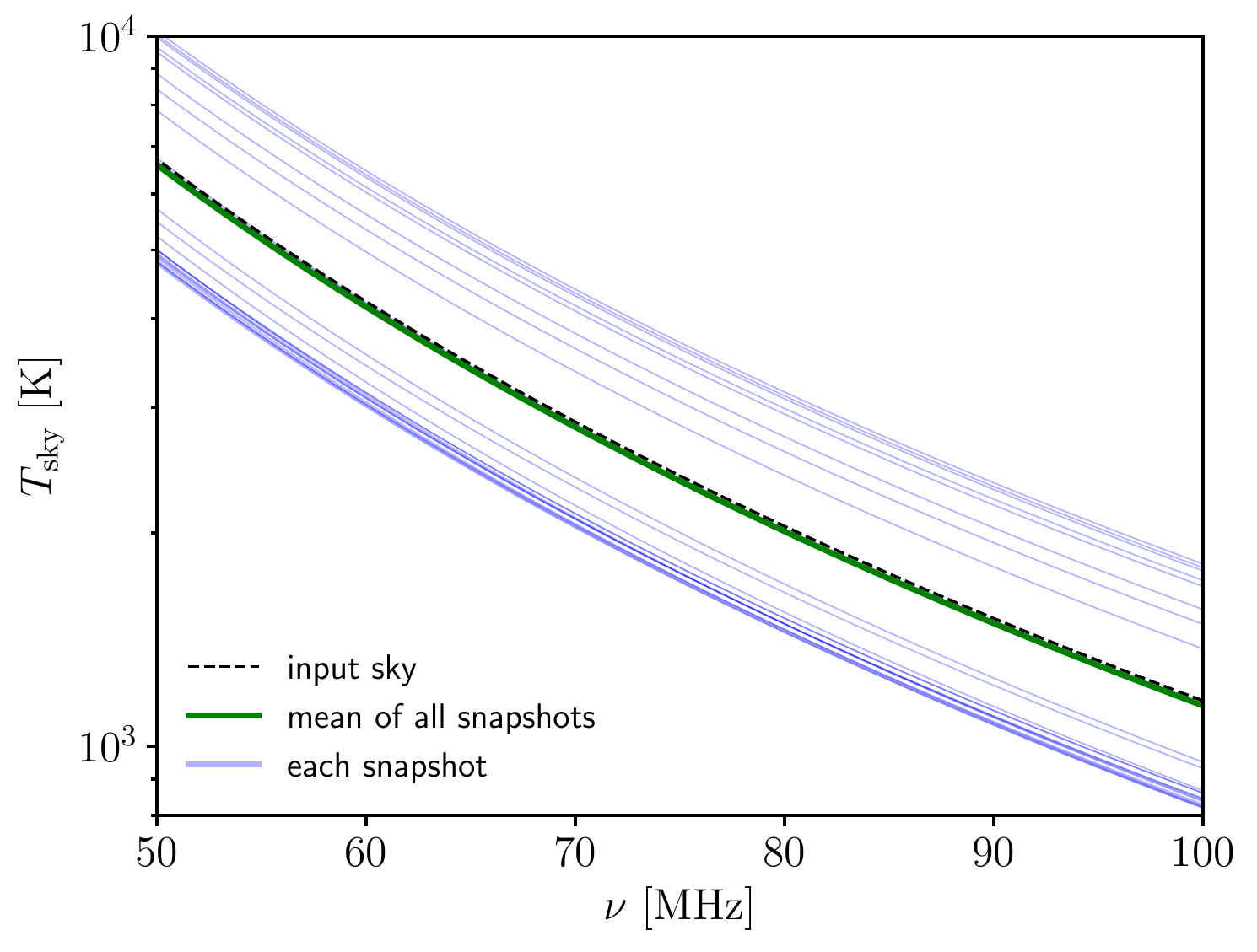}}
\subfigure{\includegraphics[width=0.4\textwidth]{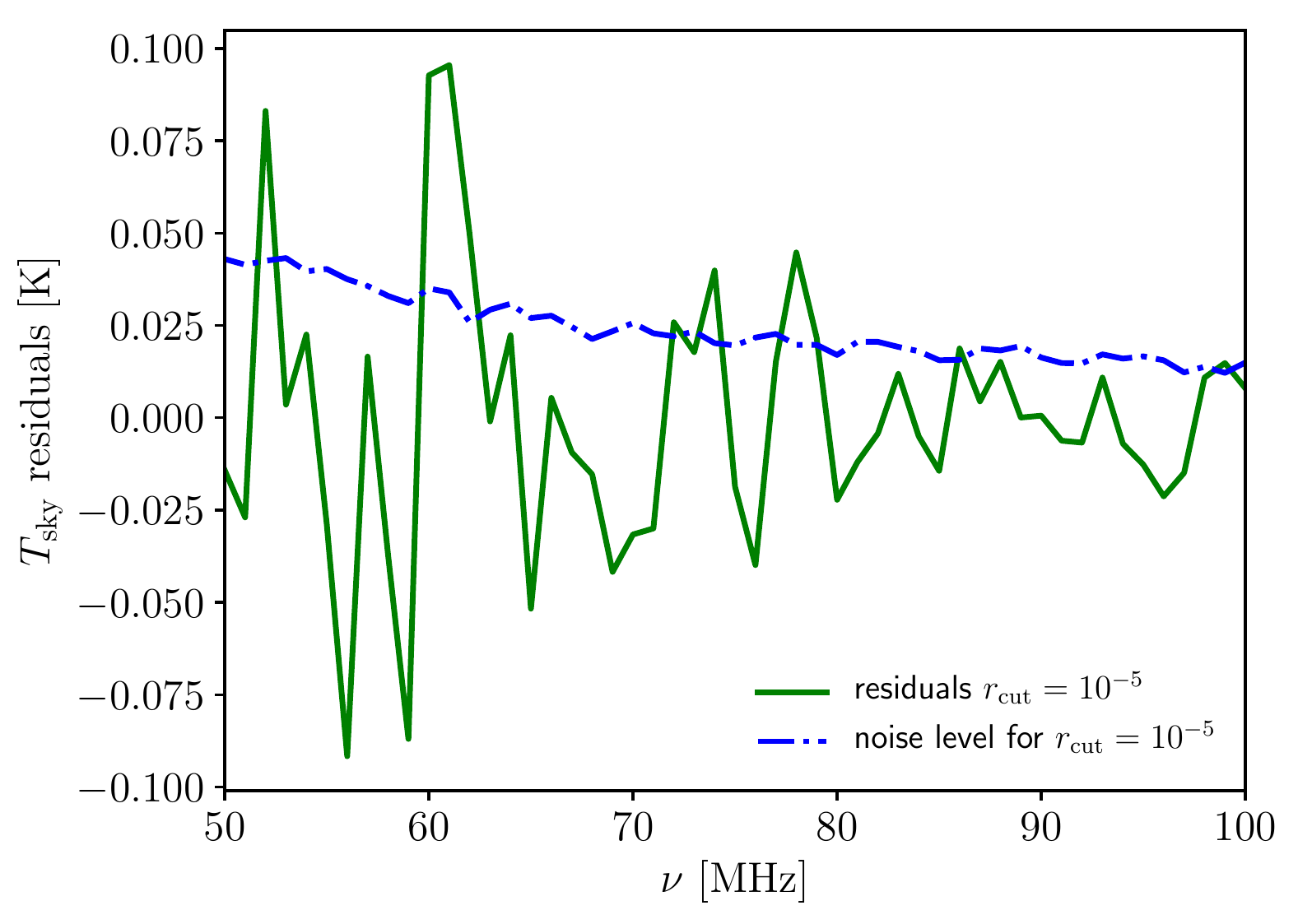}}
 \caption{Same to Fig. \ref{fig:T_recovered_ground2D}, however here it is for a ground-based 3D array.
}
\label{fig:T_recovered_ground3D}
}
\end{figure}

In Fig. \ref{fig:T_recovered_ground3D} we plot the recovered global spectrum by this ground-based 3D array, and the residuals after removing the fitted smooth component from the recovered spectrum without noise and 21 cm signal. The residual fluctuations are comparable with the noise level.

\begin{figure}
\centering{
\subfigure{\includegraphics[width=0.45\textwidth]{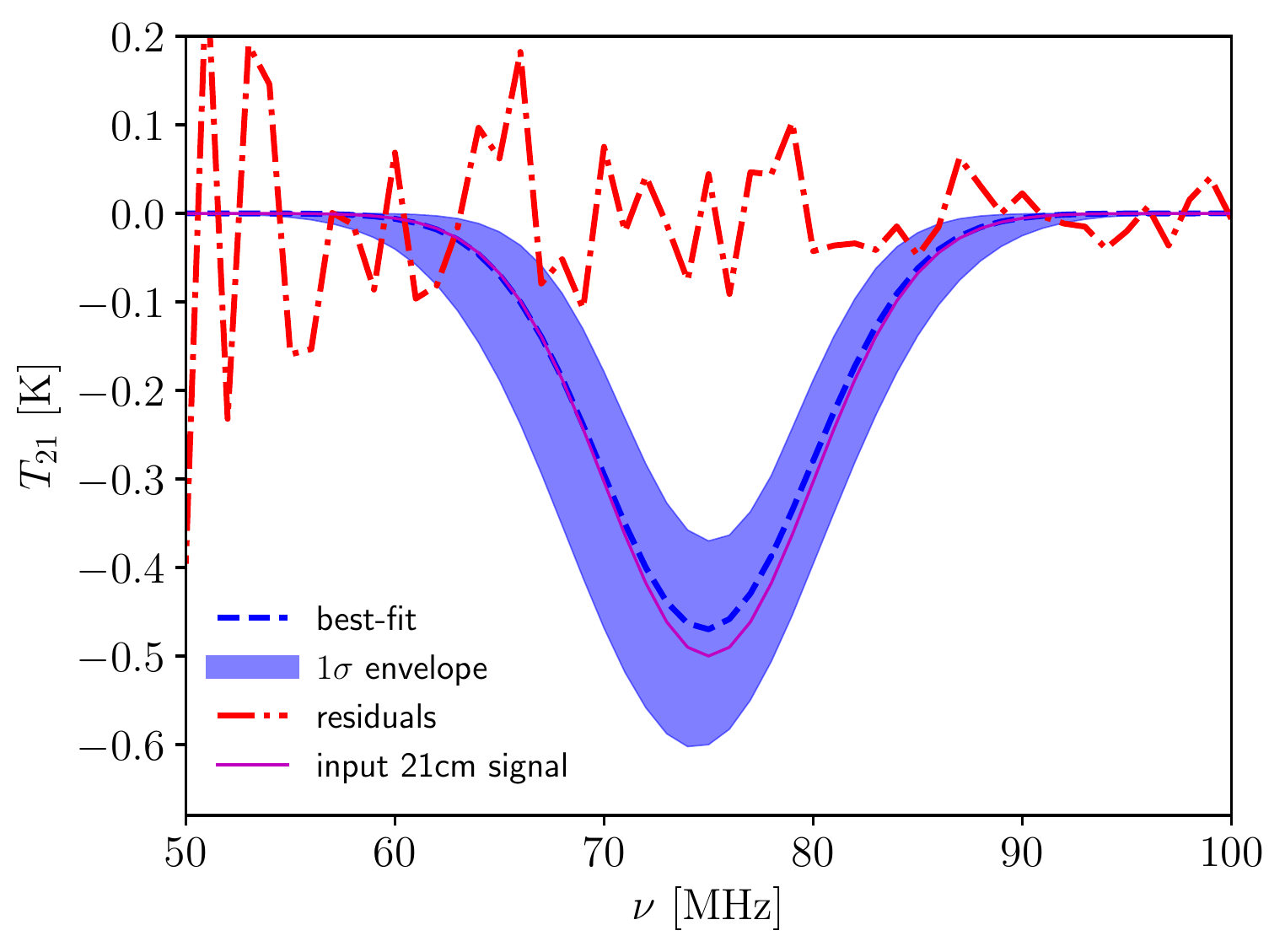}}
\caption{Same as Fig. \ref{fig:T_21cm_ground2D}, but for a ground-based 3D array.    
}
\label{fig:T_21cm_ground3D}
}
\end{figure}

In Fig. \ref{fig:T_21cm_ground3D} we plot the recovered 21 cm signal for this ground-based 3D array.
The marginalized 21 cm parameters are: $A=-0.48_{-0.05}^{+0.04}$ K, $\nu_{21}=74.8_{-0.3}^{+0.3}$ MHz and $\sigma_{21}=5.2_{-0.4}^{+0.4}$ MHz.
Here it seems that the ground-based 3D array does not perform better than the 2D array. We suspect this is because for this array-in-a-well, where the antennas are all located on a cone surface, it is not a real 3D distribution. 
Moreover, since all the beam always points to the local zenith, due to the shielding effect, at each snapshot it only observes 1/4 of the full sky (a ground-based 2D array can observe 1/2 of the full sky). As the Earth rotates, the observed part of the sky changes at different snapshots, and we solve the equations for baselines formed at each snapshot. 
For the array in the well, some sky regions may be never measured even though the Earth rotates. Also, this is different from a real 3D distribution.
A real 3D baseline distribution can be generated by antennas in space, as we will discuss next.

\subsection{A space array}\label{sec:space3D}

To overcome the shielding and/or the reflection effects, the best way is to build an array in space, at a location far from any celestial objects, for example at the Sun-Earth Lagrange point L2. In this case, if we can treat the radio sky as constant, it is not necessary to form all baselines at the same time, instead a 3D baseline distribution can be obtained by combining baselines formed at different times, through the motion of the array. One can then produce a large number of baselines with a smaller number of antennas. 

Ignoring technical details, we assume to build a cross-shape array with dipole antennas. As the baselines needed for the global spectrum measurement is relatively short, instead of individual satellites, we consider an array formed by rigid connection, which can be realized by extending a folded structure. Suppose the structure is in the form of a cross, with 6 antennas along $x$-axis and 6 along $y$-axis of the array. Their locations are randomly assigned, however the physical distance between any neighboring pairs must be $>2$ m. All of the dipole antennas are assumed to be in the linear polarization along the $x$-axis.

The array is assumed to rotate slowly about its comoving $z$-axis with rate 0.65$^\circ$ per day.  This comoving $z$-axis also rotates about the rest-frame $z$-axis with a slower rate 0.11$^\circ$ per day. The included angle between the comoving and the rest-frame $z$-axis is 30$^\circ$. Because of orbital precession, it is then possible to form many independent baselines.

\begin{figure}
\centering{
\subfigure{\includegraphics[width=0.45\textwidth]{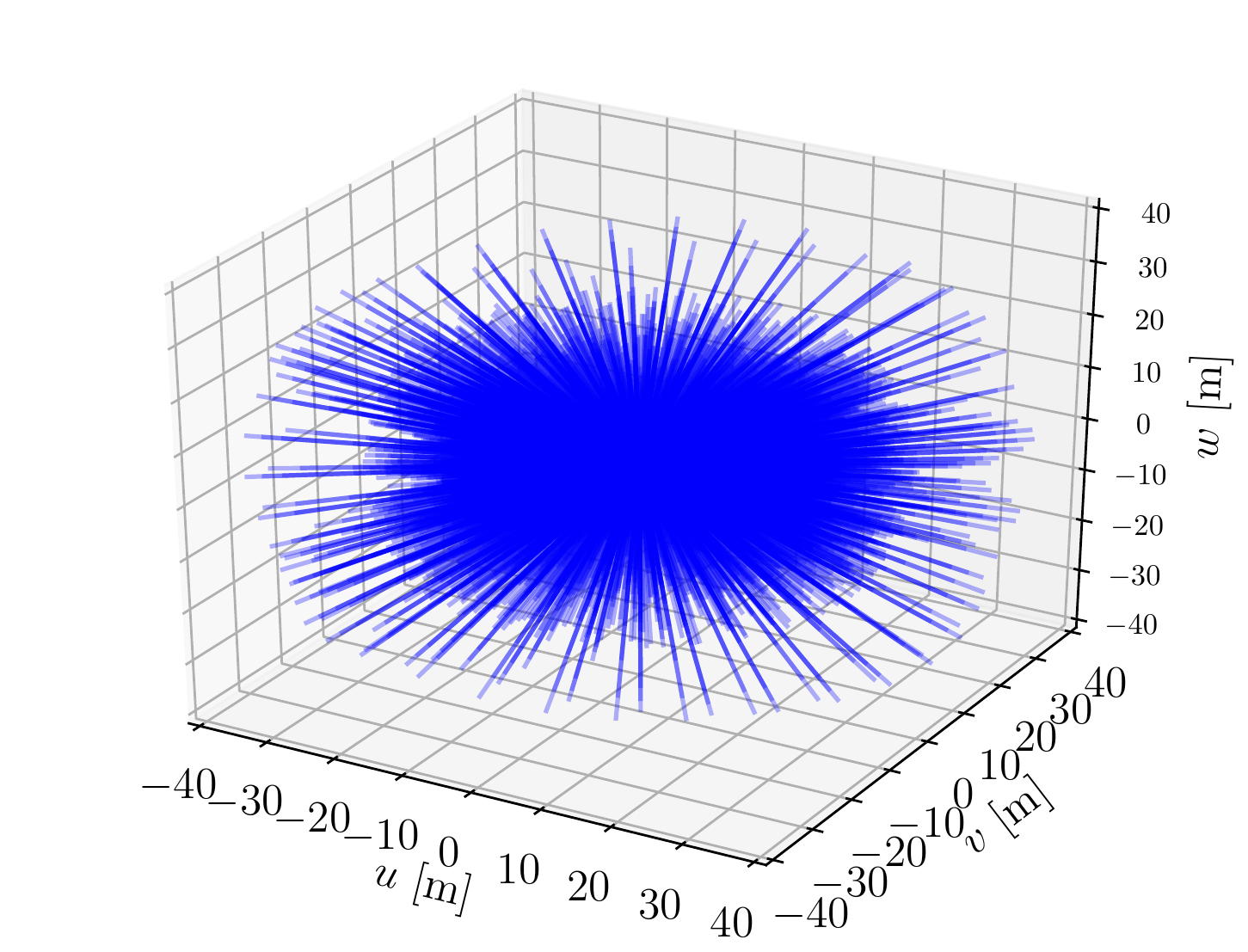}}
\caption{The baselines of the cross-shape space interferometer array obtained after 3 years.
}
\label{fig:baselines_space3D_cross}
}
\end{figure}

We take 100 snapshots in 3 years and finally get 6600 baselines. Each snapshot has integration time 262.8 hour.  Again we only choose baselines with $\lambda < b<10\lambda$ at each  wavelength. All baselines are shown in Fig. \ref{fig:baselines_space3D_cross}.

\begin{figure}
\centering{
\subfigure{\includegraphics[width=0.4\textwidth]{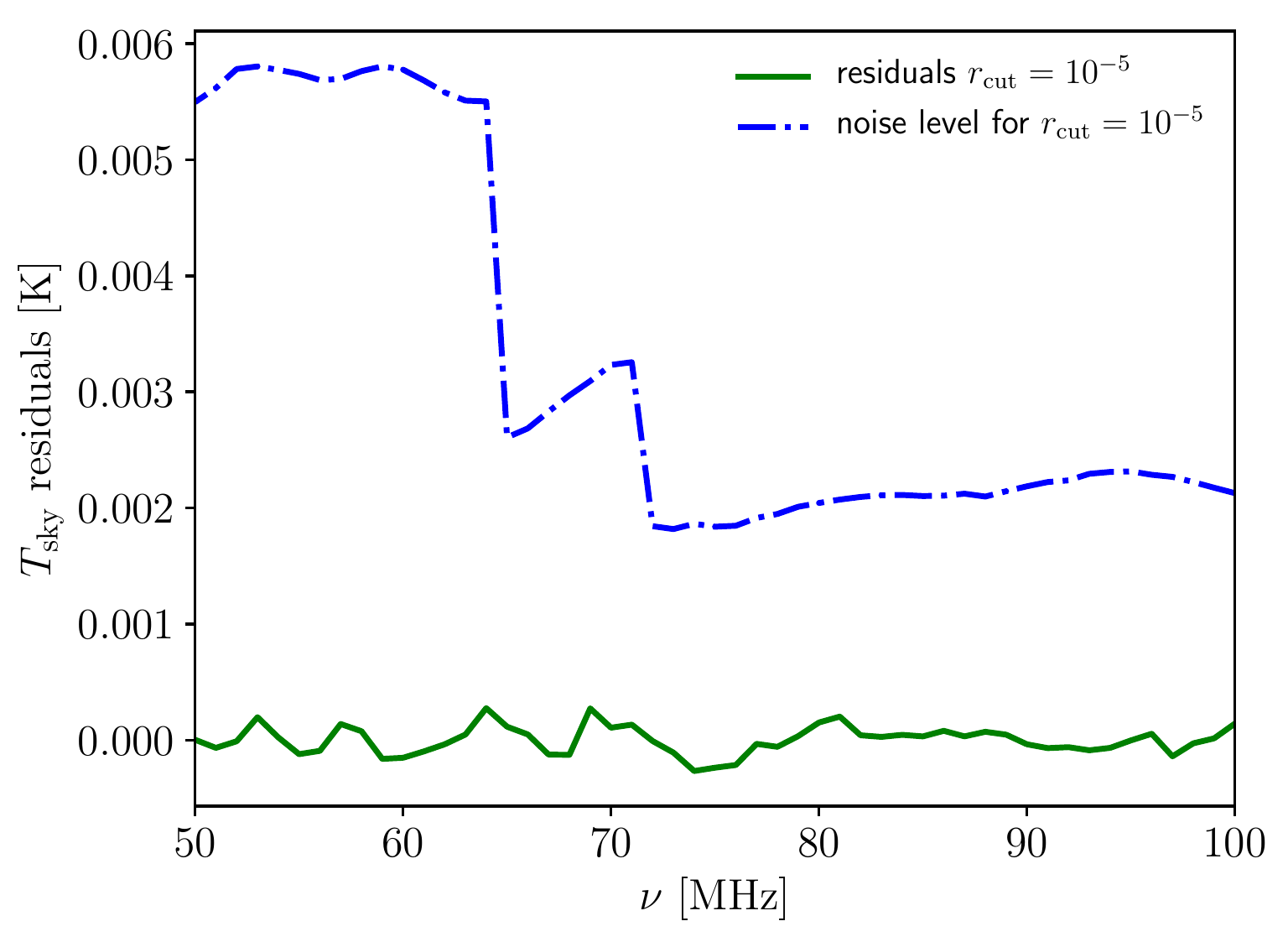}}
 \caption{
Same to bottom panel of Fig. \ref{fig:T_recovered_ground2D}, however here it is for a cross-shaped space array. We see that the residual fluctuations in the recovered global spectrum is almost negligible.
}
\label{fig:T_residuals_space3D}
}
\end{figure}

In Fig. \ref{fig:T_residuals_space3D} we show the residuals after removing the best-fit smooth component from the recovered global spectrum without noise and 21 cm signal. 
We see that unlike the ground-based arrays, here the residual fluctuations are negligible compared with the noise level. Moreover, here the residual fluctuations are not sensitive to the choice of $r_{\rm cut}$.
We also find that the noise as a function of frequency is incontinuous. This is because although we have thousands of baselines, they are generated by just 12 antennas through orbit precession.
Orbit precession changes the orientations of baselines but does not change their lengths. Therefore at each frequency there are only several tens of different baseline lengths. Their lengths distribution must be sparse and inhomogeneous in the range $\lambda <b<10\lambda$. In this case,  the noise level is somewhat opportunistic. 
At frequencies where the shortest baselines are closer to $\lambda$, the noise is smaller.  
For example, at 65 MHz the shortest baselines are $\approx\lambda$, while at 64 MHz the shortest baselines are $\approx 1.5\lambda$ because the shortest baselines at 65 MHz are already shorter than $\lambda$ at 64 MHz and are discarded. So the noise jumps sharply from 65 MHz to 64 MHz. The reason for the noise jump from 72 MHz to 71 MHz is the same. Moreover, from 100 MHz to 72 MHz, the noise is roughly constant instead of increasing like the $T_{\rm sky}(\nu)$ trend. Because in this frequency range the shortest ones of baselines with $\lambda <b<10\lambda$ are the same, hence $b_{\rm min}/\lambda$ decreases gradually with decreasing frequency. It compensates for the increasing of $T_{\rm sky}$, as a result the noise is roughly constant. 
Nevertheless, since the amplitude of the noise is already small enough, no matter whether it is continuous or not, the influence on  the 21 cm signal is small. The smoothness of the noise can be improved by increasing the number of antennas.

\begin{figure}
\centering{
\subfigure{\includegraphics[width=0.45\textwidth]{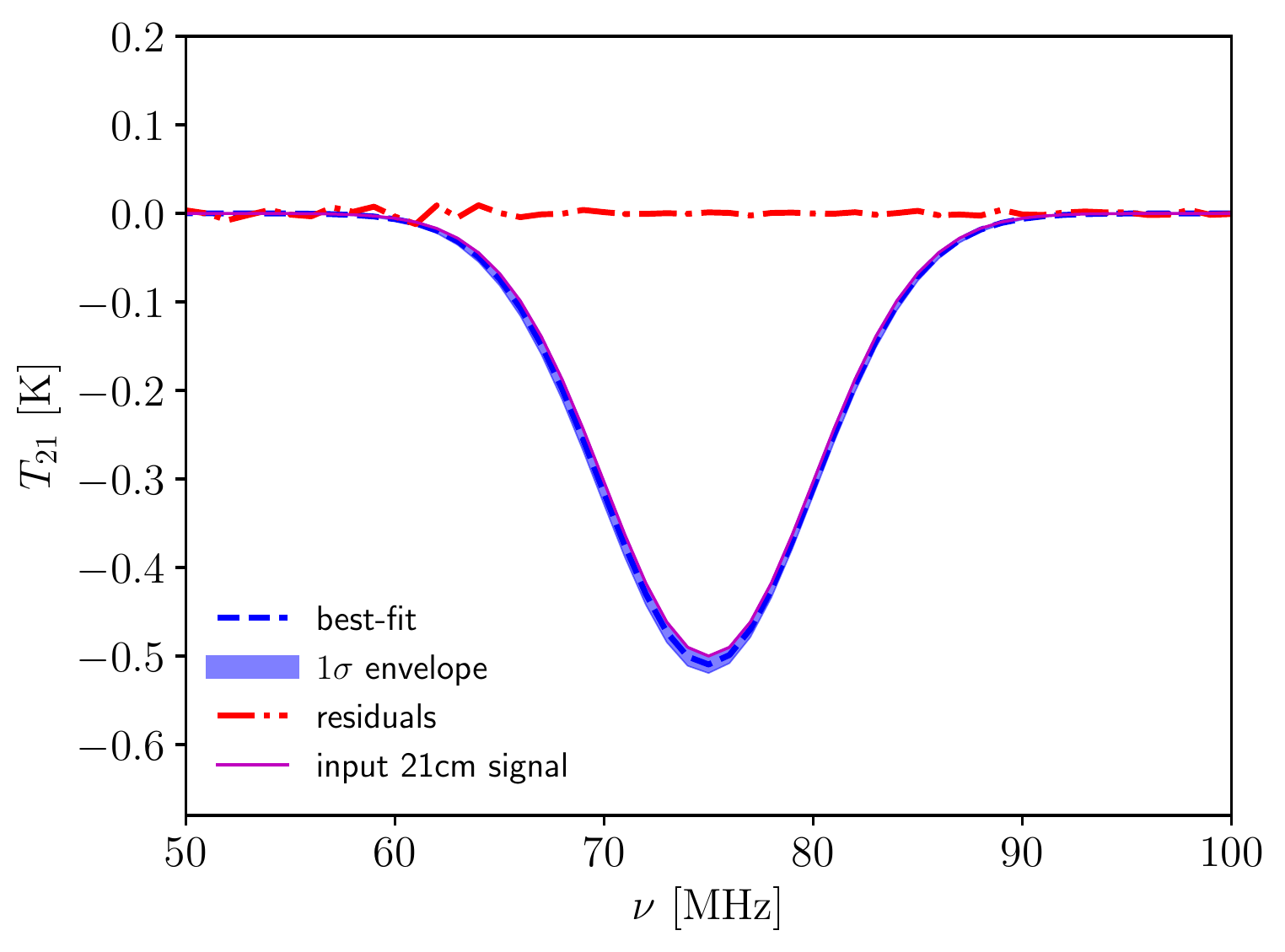}}
\caption{Same to Fig. \ref{fig:T_21cm_ground2D}, however here it is for  cross-shaped space array.   
}
\label{fig:T_21cm_space3D}
}
\end{figure}

In Fig. \ref{fig:T_21cm_space3D} we plot the recovered 21 cm signal for this cross-shaped space array. The 21 cm signal parameters are 
$A=-0.509_{-0.004}^{+0.004}$ K, $\nu_{21}=74.96_{-0.02}^{+0.02}$ MHz and $\sigma_{21}=5.07_{-0.03}^{+0.03}$ MHz.
These results show that the space array, without the blocking of sky, could recover the global spectrum of the sky better, though realizing such array would be much harder than ground arrays.

\subsection{Compared with previous works}

As noted in the {\it Introduction}, in this paper we considered general solution of the monopole and multipole spectrum from interferometer cross-correlations. In \citet{Presley2015}, instead,
an estimator for the global sky temperature was proposed,
\begin{equation}
\hat{T}_{\rm sky}= 
\frac{ \sum_j\left[ \int d \Omega(\hat{\boldsymbol{n}}) B_\nu(\hat{\boldsymbol{n}})  e^{-2 \pi i \frac{\boldsymbol{b}_j}{\lambda} \cdot \hat{\boldsymbol{n}}} \right]  V(\boldsymbol{b}_j)  } {\sum_k |\int d \Omega(\hat{\boldsymbol{n}}) B_\nu(\hat{\boldsymbol{n}})  e^{-2 \pi i \frac{\boldsymbol{b}_k}{\lambda} \cdot \hat{\boldsymbol{n}}}|^2}.
\label{eq:estimator_P15}
\end{equation}
Here the anisotropy of the 21 cm signal is very small so it could be neglected, but the foreground anisotropy may contribute to it. In this approach a direct measurement is attempted, and it only requires a handful of baselines, so a regular grid array with many redundant baselines can be used, while in our approach we solve for the monopole and higher-order spherical harmonic coefficients simultaneously, and  to obtain the accurate global spectrum we generally need $10^2 \sim 10^4 $ independent baselines, which can be realized with an array of antennas at randomized locations.

\begin{figure}
\centering{
\subfigure{\includegraphics[width=0.4\textwidth]{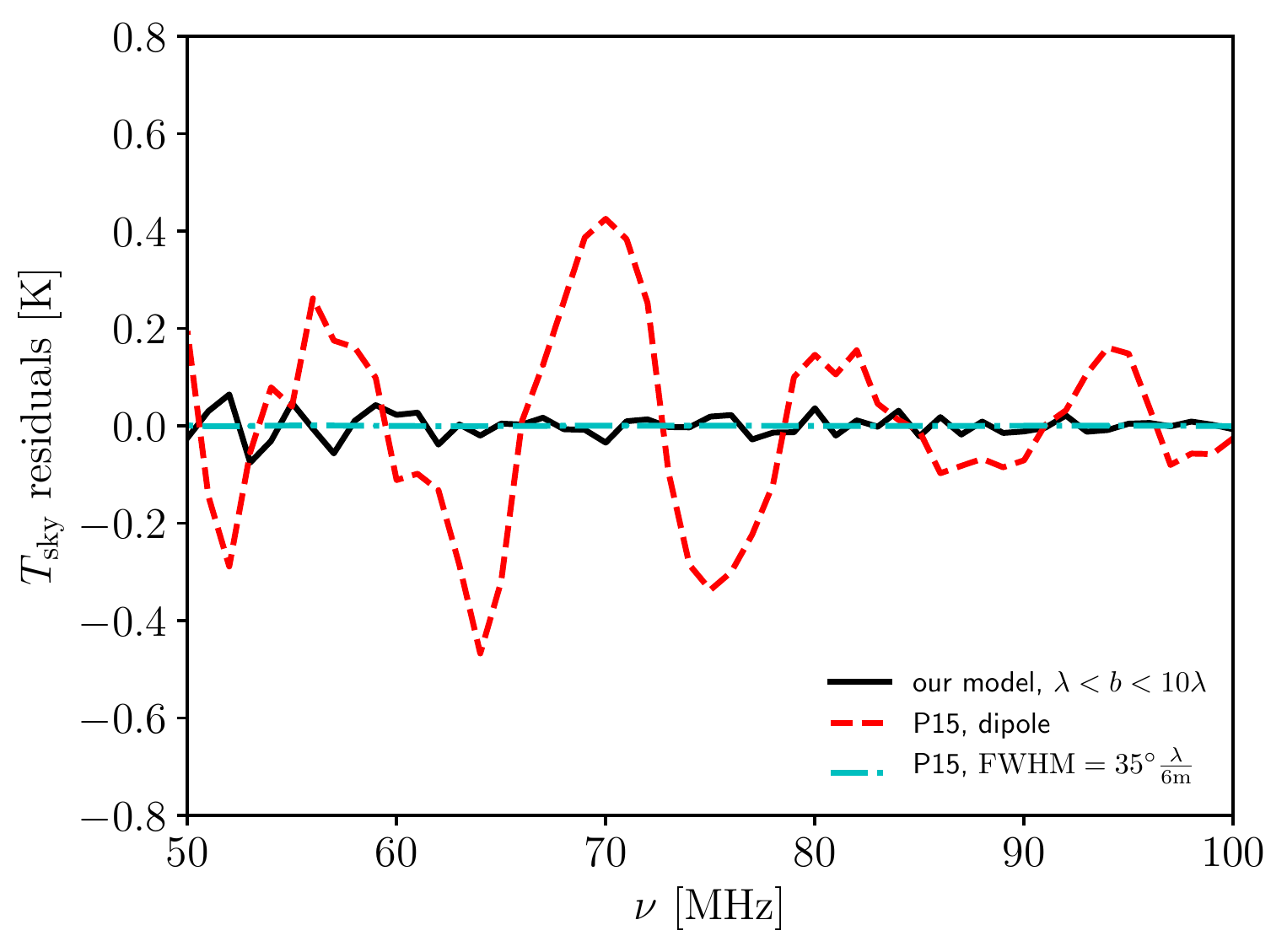}}
 \caption{
 The residuals after we remove the best-fit 5-order polynomial in the recovered foreground global spectrum. The foreground does not contain noise so the residual fluctuations are purely induced by the array configuration. We check that using 7-order polynomial will just slightly reduce the residuals.
}
\label{fig:T_residuals_ground2D_P15}
}
\end{figure}

For Eq. (\ref{eq:estimator_P15}) to work well, there is an optimal baseline length that depends on the beam size. 
\citet{Presley2015} assumed that the beam size is proportional to the wavelength, so that within the observational frequency band, a fixed physical length will be optimal. However, if the beam size is not 
exactly proportional to the wavelength, 
it may generate extra fluctuations on the global spectrum. We test two cases for our ground-based 2D array about this estimator: a) if we use a Gaussian beam with ${\rm FWHM}=35^\circ\frac{\lambda}{\rm 6m}$;
b) if we directly apply  it to dipole antennas. 
Same to \citet{Presley2015}, the Gaussian beam is multiplied by a cos term so it gradually drops to zero at the horizon.
In this test we keep all baselines, even those shorter than the wavelength. 
The residuals after removing the best-fit 5-order polynomial are shown in Fig. \ref{fig:T_residuals_ground2D_P15}. We see that the estimator Eq. (\ref{eq:estimator_P15})
performs well for the ideal Gaussian beam (beam width exactly proportional to the wavelength), but have large residues for the dipole beam.

\subsection{The dependence on beam form}\label{sec:beam_dependence}

\begin{figure}
\centering{
\subfigure{\includegraphics[width=0.4\textwidth]{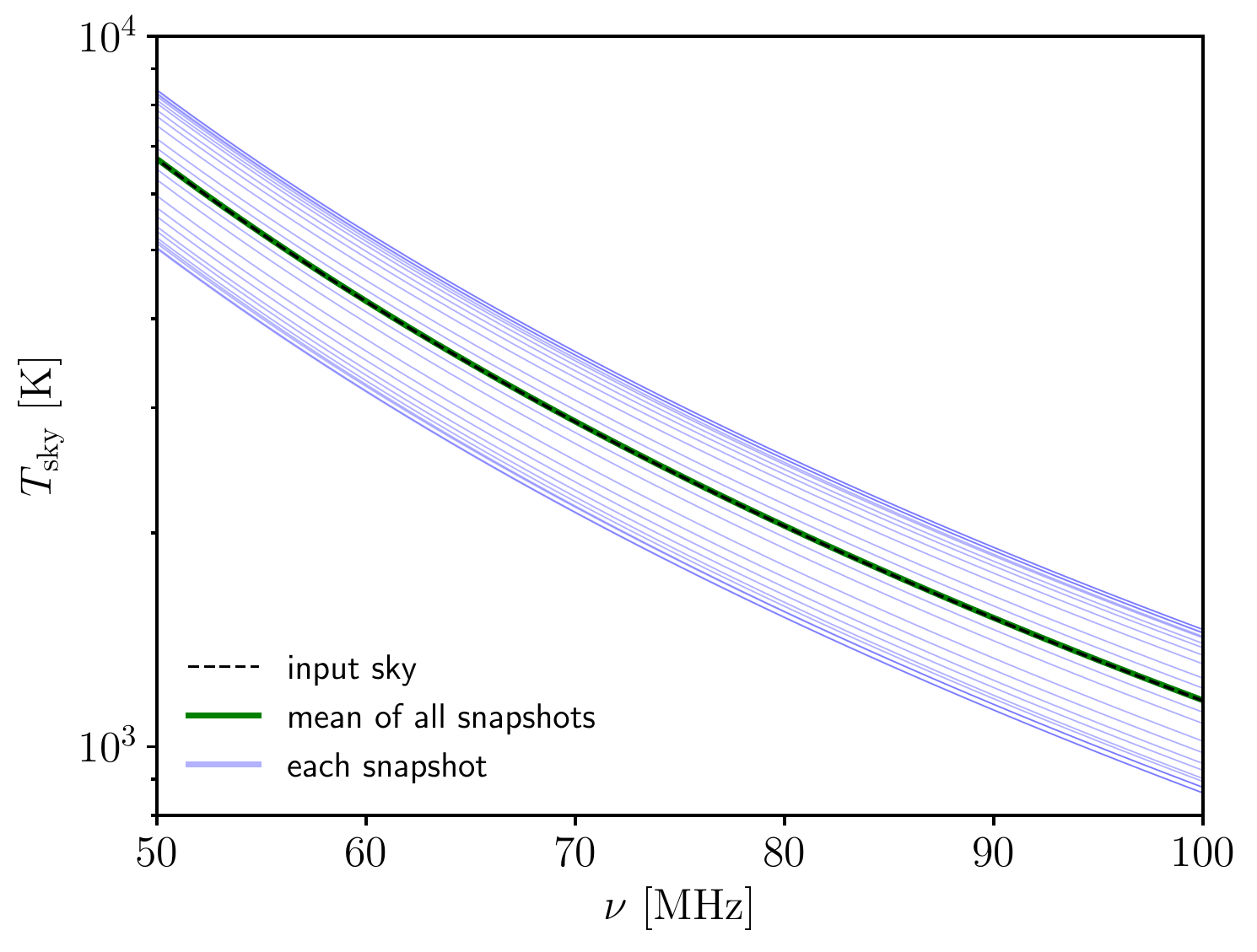}}
\subfigure{\includegraphics[width=0.4\textwidth]{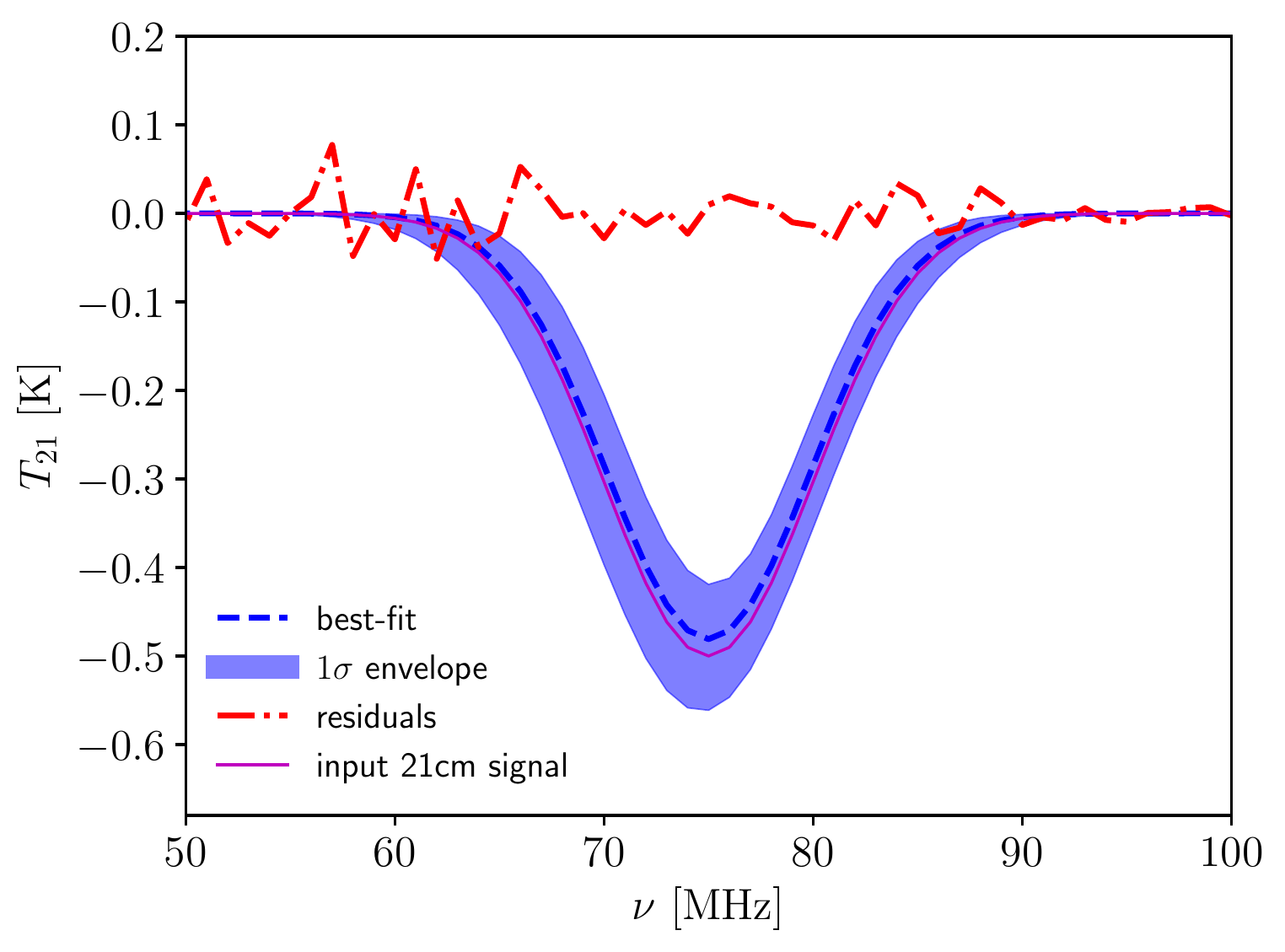}}
 \caption{{\it Top:} The recovered global sky temperature for ground-based 2D array in Sec. \ref{sec:ground2D}, however the beam is Gaussian form.
 {\it Bottom:} The extracted 21 cm signal.
}
\label{fig:fig17}
}
\end{figure}

\begin{figure}
\centering{
\subfigure{\includegraphics[width=0.4\textwidth]{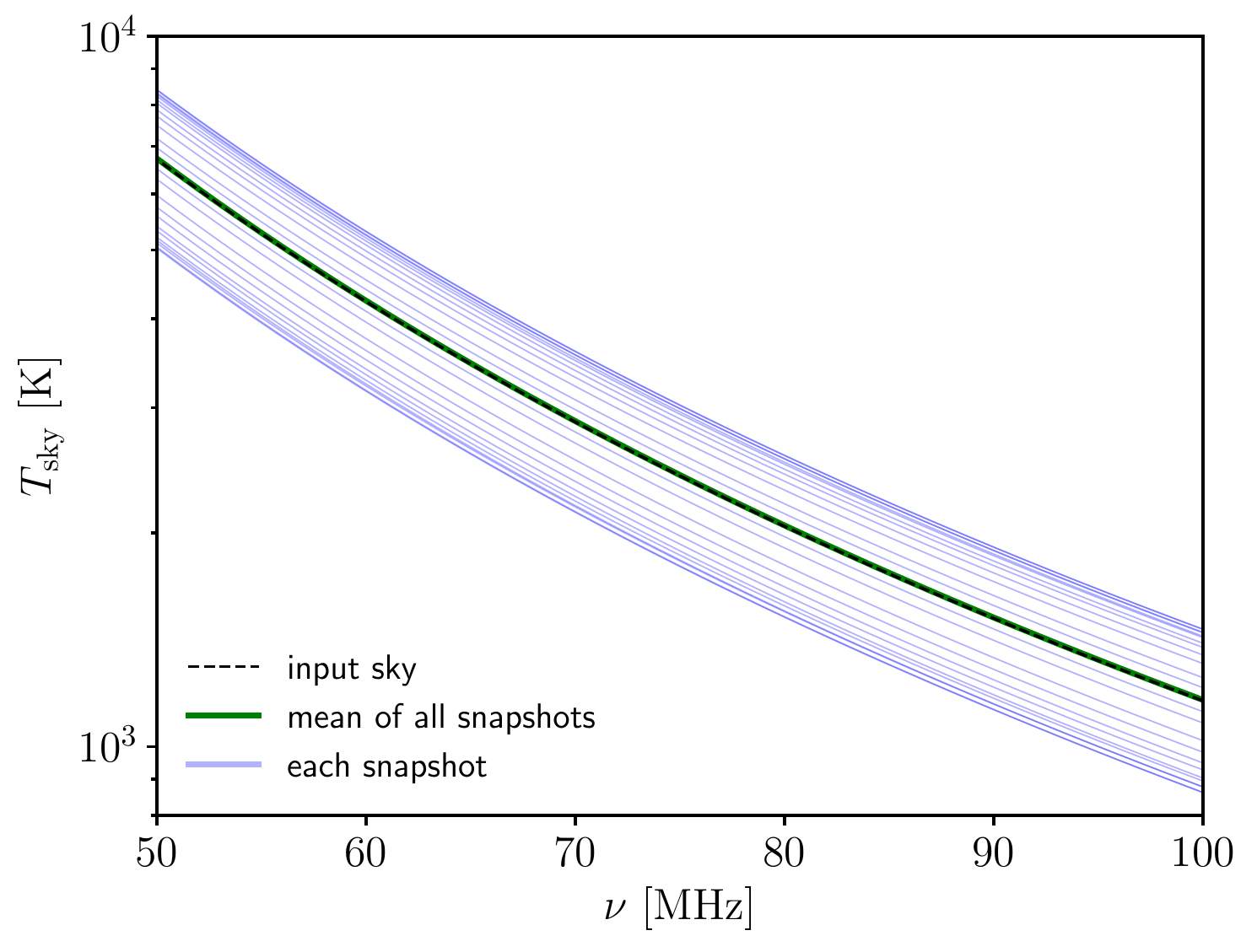}}
\subfigure{\includegraphics[width=0.4\textwidth]{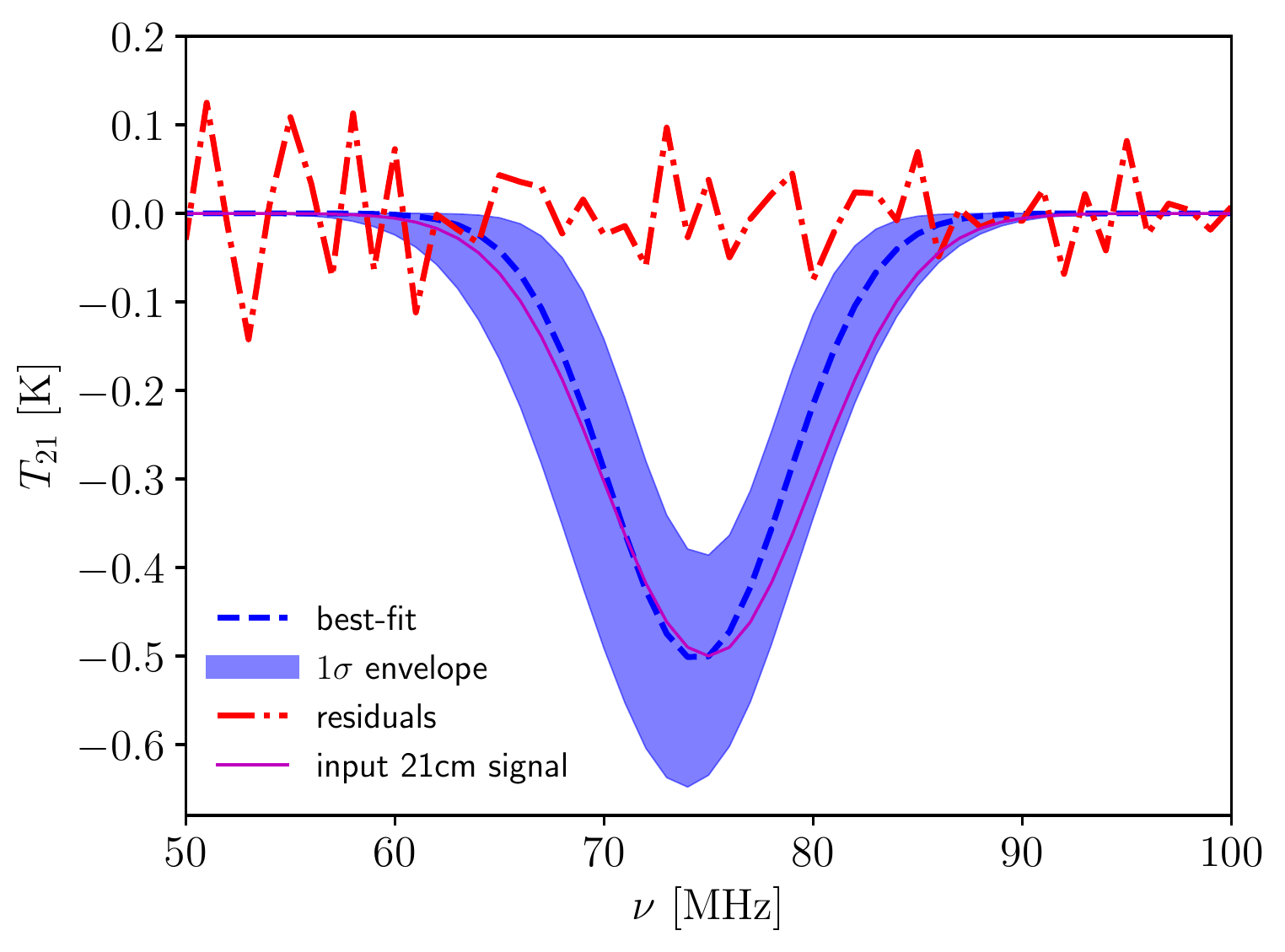}}
 \caption{Same to Fig. \ref{fig:fig17}, however here the beam is $\cos^2(f\theta)$ .
}
\label{fig:fig18}
}
\end{figure}

We have adopted dipole antenna for our simulations, since it is the simplest (both in theory and in technique) antenna.
However, we do not specify any particular properties for the beam, 
so our methods are actually applicable in broad range of conditions. 
For the ground-based 2D array in Sec. \ref{sec:ground2D}, we show the recovered global sky temperature and the extracted 21 cm signal for other two kinds of beams.
Fig. \ref{fig:fig17} is for a Gaussian beam
$B(\theta)=\exp\left(-\frac{\theta^2}{2\sigma^2_{\rm B}} \right)$,
where $\theta$ is the angular distance to the local zenith, and we adopt $\sigma_{\rm B}=30^\circ$.
For the reason mentioned in Sec. \ref{sec:beam}, the recovered sky temperature has underestimated bias $\sim0.1\%$.
However, if the Gaussian beam drops to zero gradually toward the horizon, i.e. 
$B(\theta)=\exp\left(-\frac{\theta^2}{2\sigma^2_B} \right) \cos(\theta)$ \citep{Presley2015}, then the underestimated bias increases to $\sim 4\%$. 
In this case, to reduce the bias one can solve the visibility equations of 3D baselines formed via Earth rotation, as we mentioned in Sec. \ref{sec:beam}.
Fig. \ref{fig:fig18} is for a beam 
$B(\theta)=\cos^2(f\theta)$
and we adopt $f=0.8$.
We see that the global sky temperature and 21 cm signal are still recovered well. It means that we do not need to specify a particular beam form when constructing the array. 
In practice, the beam is provided by direct measurements using the calibration source, or from the parameterization fitted by measurements.

We have ignored the uncertainties of the beam.  Knowing the beam perfectly is a challenge for all interferometers that aim to detect 21 cm signal. We find that, suppose there is Gaussian beam constructed from 8000 measurement points each has relative error $\sim5\%$, then the beam spread $\sigma_B$ can be constrained to $\sim0.01\%$ level. If at each frequency the beam errors are independent, they lead to $\sim 0.04\%$ level extra error fluctuations on the recovered foreground global spectrum.
Such fluctuations are comparable to the 21 cm signal. So to measure the 21 cm signal perfectly, generally it requires the beam uncertainties $\ll \sim 5\%$.

\subsection{The cross-talk effect}

So far we have not yet considered the cross-talk between antenna pairs. Cross-talk can origin from the internal noise of one antenna that leaks into another one; and the sky signal scattered by one antenna that received by another one \citep{Thekkeppattu2022}.
\citet{2016ApJ...826..116V} pointed out that cross-talk is inevitable even for the ideal interferometric setup, because of the scattering of sky radiation by each of the paired antennas. The attempt to suppress the cross-talk by reducing the antenna size and increasing the pair separation will also reduce the sensitivity to sky signal. If the origin of cross-talk is well-known and well-calibrated, in the visibility it can be modeled simultaneously with the cross-correlation.
In this subsection, we show the feasibility of solving the global spectrum in the presence of cross-talk in our method.

According to \citet{Thekkeppattu2022}, the voltages of the two antennas are:
\begin{align}
e_1&=e_{\rm 1,sky}+e_{\rm 1,RX}+f_c(\nu,\boldsymbol{b})(e_{\rm 2, sky}+e_{\rm 2,RX} ) \nonumber \\
e_2&=e_{\rm 2,sky}+e_{\rm 2,RX}+f_c(\nu,\boldsymbol{b})(e_{\rm 1, sky}+e_{\rm 1,RX} ), 
\end{align}
where $e_{\rm sky}$ is from the sky signal, $e_{\rm RX}$ is from the internal noise, $f_c$ is the coefficient describes the strength of cross-talk. Suppose the two antennas are identical, the visibility in the presence of cross-talk is
\begin{align}
V'_{12}&=\mean{e_1e^*_2 }  \nonumber \\
&=(V_{12}+f^*_cT^\otimes_{\rm sky}+f_c T^\otimes_{\rm sky}+f_cf^*_c V^*_{12})+(f_c^* T_{\rm RX}+f_c T_{\rm RX} ), 
\label{eq:V_12_p}
\end{align}
where $V_{12}$ is the visibility in the absence of cross-talk; $T^\otimes_{\rm sky}$ is the convolution of sky temperature and beam; $T_{\rm RX}$ is the internal noise temperature. Obviously, now the measured visibility contains not only the cross-correlation, but also partial contributions from the auto-correlation of the two antennas.

We can also expand Eq. (\ref{eq:V_12_p}) like Eq. (\ref{eq:V2}), and finally write the equations for many baselines 
\begin{align}
\boldsymbol{V'}&=[\boldsymbol{Q}+ (f_c+f_c^*) \boldsymbol{R} +f_cf_c^*\boldsymbol{P} ]\boldsymbol{a}+\boldsymbol{V'}_{\rm N} \nonumber \\
&=\boldsymbol{Q'} \boldsymbol{a}+\boldsymbol{V'}_{\rm N},
\label{eq:V_p}
\end{align}
where
\begin{equation}
R_{l,j}^m=\left(\int d \Omega(\hat{\boldsymbol{n}}) B_\nu(\hat{\boldsymbol{n}}) Y_l^m(\hat{\boldsymbol{n}} )   \right),
\end{equation}
and
\begin{equation}
P_{l,j}^m=\left(\int d \Omega(\hat{\boldsymbol{n}}) B(\hat{\boldsymbol{n}}) Y_l^m(\hat{\boldsymbol{n}} )  e^{2 \pi i \frac{\boldsymbol{b}_j}{\lambda} \cdot \hat{\boldsymbol{n}}}\right).
\end{equation}
Similar to Eq. (\ref{eq:V_tilde}), we re-write a new equation 
\begin{equation}
\tilde{\boldsymbol{V}}'=[\tilde{\boldsymbol{Q}} + 2{\rm Re}(f_c) \tilde{\boldsymbol{R}} + |f_c|^2 \tilde{\boldsymbol{P}}] \tilde{a},
\label{eq:V_p_tilde}
\end{equation}
where $\tilde{\boldsymbol{R}}$ and $\tilde{\boldsymbol{P}}$ are constructed similar to $\tilde{\boldsymbol{Q}}$, except that the $Q_{l,j}^m$s in Eq. (\ref{eq:ABCD}) are replaced with $R_{l,j}^m$s or  $P_{l,j}^m$s. 
Suppose the cross-talk coefficient $f_c(\nu,\boldsymbol{b})$ is known, 
we can still solve Eq. (\ref{eq:V_p_tilde}) using the method same to Eq. (\ref{eq:V_tilde}).

\begin{figure}
\centering{
\subfigure{\includegraphics[width=0.45\textwidth]{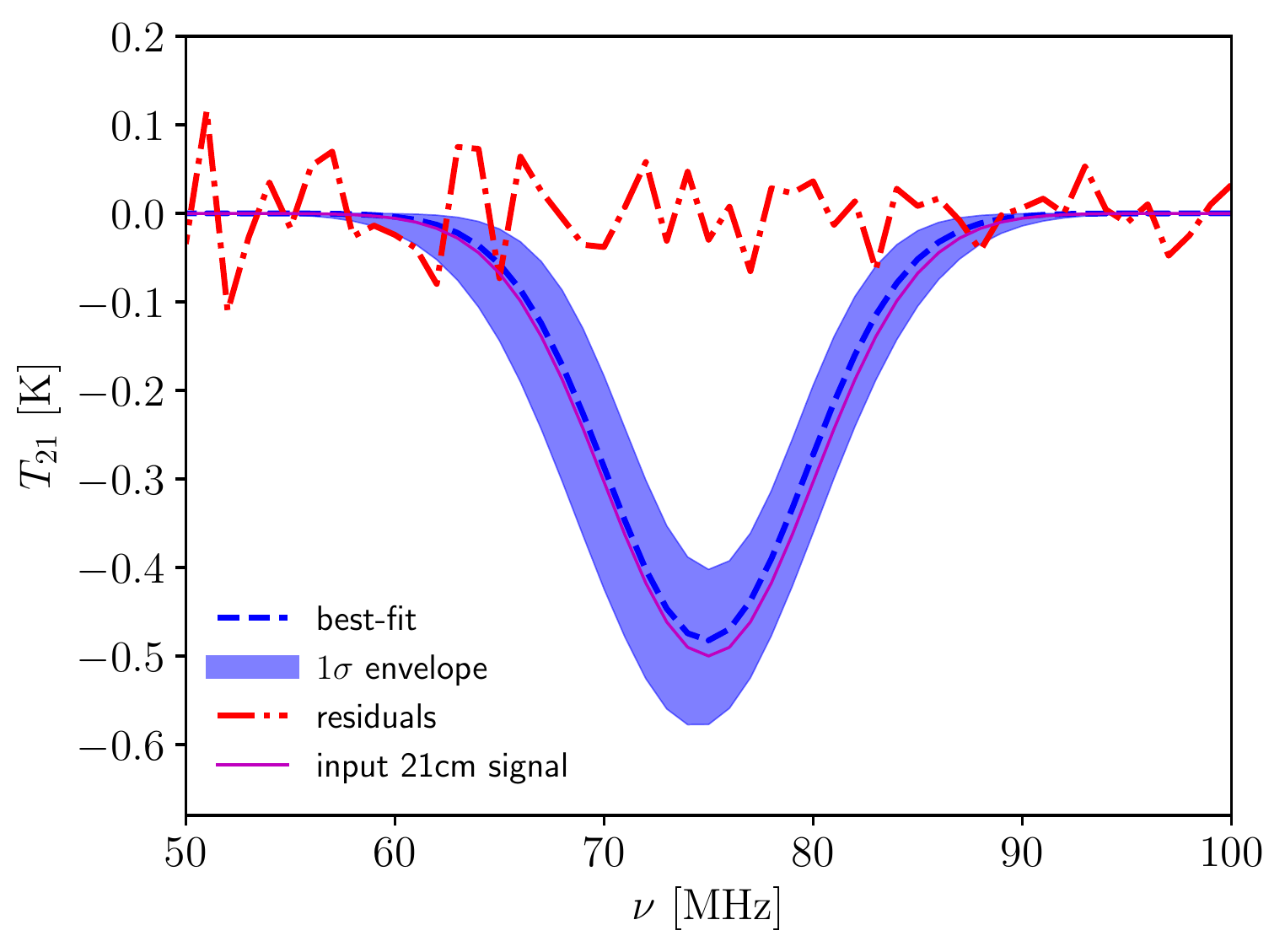}}
\caption{Same as Fig. \ref{fig:T_21cm_ground2D}, however here the 21 cm signal is solved in the presence of cross-talk.    
}
\label{fig:T_21cm_ground2D_cross-talk}
}
\end{figure}

We simply assume that the cross-talk has a smooth component plus a frequency-dependent ripple. The ripple is generated by resonant reflections (e.g. \citealt{Li2021RAA,Sun2022RAA}). The cross-talk amplitude is inversely proportional to the baseline length.
It writes
\begin{equation}
f_c(\nu,\boldsymbol{b})=0.01\left(\frac{b}{\rm 6m}\right)^{-1}[1+e^{2\pi i \tau_b \nu} ], 
\end{equation}
where $\tau_b=2b/c$, $c$ is the speed-of-light.
For our ground-based 2D array in Sec. \ref{sec:ground2D}, we show the recovered 21 cm global spectrum in Fig. \ref{fig:T_21cm_ground2D_cross-talk}.  Not surprisingly, the 21 cm signal is still recovered well, and the results are close to the results solved purely from cross-correlation.

So we can still solve the global sky temperature even the cross-talk is taken into account. Just, in this case the solved temperature contains some information (and potential interloping) from the auto-correlation. This is enough since the current paper is a theoretical investigation.  
Moreover, when solve the Eq. (\ref{eq:V_p}) we do not require  
to specify a particular form for $f_c$. Therefore this method is applicable to any cross-talk form, as long as it is known. 
In practice however, the cross-talk can be more complicated and the methods to calibrate the $f_c(\nu,\boldsymbol{b} )$ depend on details of the constructed array. That is beyond the scope of this paper.

\section{Conclusions}\label{sec:conclusion}

We investigated the feasibility of recovering the global sky temperature from visibilities measured by  interferometer array with baselines $>\lambda$,  and of extracting  21 cm global spectrum from the recovered temperature.  We found that:

\begin{itemize}

\item  The global sky temperature of both the foreground and the 21 cm signal can be recovered from the visibilities measured by interferometers with up to thousands of baselines.
The precision depends on the noise and the completeness of baseline distribution. 
The 3D baseline distributions have much better performance than the 2D baseline distribution. For 3D baselines, the global sky temperature can be recovered even when the shortest baselines are much longer than the wavelength.

\item  We made simulations for ground-based 2D, ground-based 3D, and space array configurations. 
For ground-based interferometer, 
because the hemisphere under the horizon is blocked by the Earth, and as Earth rotates the hemisphere in field-of-view changes gradually,  we use the instantaneous $uv$-coverage to recover the global sky temperature. 
For space array however, it can be located at, for example, the Sun-Earth L2 point, so that it receives the all-sky radiation simultaneously. Through orbit procession, the 3D baseline distribution can then be obtained by combining baselines at different time.

\item In addition to noise, in the recovered global spectrum there are fluctuations caused by non-perfect baseline distribution and cutoff loss, particularly for the ground-based array. However we have checked that these fluctuations can be well controlled (below or comparable with the noise) if  reasonable array configuration is chosen. In all cases one can obtain good global sky temperature and extract the correct 21 cm signal from it, as long as the integration time is sufficiently long. 
For example, if the input test 21 cm signal has amplitude $-0.5$ K, 
 for a ground-based 2D array with 400 dipole antennas and $100$ days observation time, 
 we obtain $-0.49_{-0.05}^{+0.04}$ K; a ground-based 3D array with same number of dipole 
antennas and observation time we obtain $-0.48_{-0.05}^{+0.04}$ K.
For a cross-shaped space array with 12 dipoles and 3 years observation time
we obtain $-0.509_{-0.004}^{+0.004}$ K.
Obviously, space array has much better performance than ground-based arrays.

\item While we mainly presented results for the dipole beam, our method does not require any particular shape for the beam form. We checked that all conclusions will not change if we use the Gaussian or $\cos^2(f\theta)$ beam forms instead.
The methods are applicable in broad range of conditions.

\end{itemize}

\section*{Acknowledgments} 
We thank the anonymous referee for the very helpful comments. 
 This work is supported by National SKA Program of China, grant No. 2020SKA0110402, the MoST-BRICS Flagship Project 2018YFE0120800, the Chinese Academy of Sciences (CAS) Strategic Priority Research Program XDA15020200, the CAS Key Instrument Grant ZDKYYQ20200008, the National Natural Science Foundation of China (NSFC) grant 11973047, 11633004, and the CAS Frontier Science Key Project QYZDJ-SSW-SLH017. 
This work used computing resources of the Astronomical Big Data Joint Research Center, co-founded by National Astronomical Observatories, Chinese Academy of Sciences and Alibaba Cloud, and the computing resources of the National Supercomputing Center in Tianjin.




\bibliography{refe}{}
\bibliographystyle{aasjournal}



\end{document}